\documentclass[pdflatex,sn-mathphys-num]{sn-jnl}

\usepackage{graphicx}%
\usepackage{multirow}%
\usepackage{amsmath,amssymb,amsfonts}%
\usepackage{amsthm}%
\usepackage{mathrsfs}%
\usepackage[title]{appendix}%
\usepackage{xcolor}%
\usepackage{textcomp}%
\usepackage{manyfoot}%
\usepackage{booktabs}%
\usepackage{algorithm}%
\usepackage{algorithmicx}%
\usepackage{algpseudocode}%
\usepackage{listings}%

\usepackage{array}
\usepackage{tabularx}
\usepackage{float}

\usepackage{booktabs}

\usepackage{tcolorbox} 
\usepackage{xcolor} 
\usepackage{enumitem} 
\newtcolorbox{hazardcard}[2][]{ 
colback=gray!3, 
colframe=gray!55, 
title=\textbf{#2}, 
fonttitle=\small, 
fontupper=\footnotesize, 
boxrule=0.4pt, 
arc=2pt, 
left=4pt, 
right=4pt, 
top=4pt, 
bottom=4pt, 
before skip=6pt, 
after skip=6pt, #1 }

\usepackage{soul}
\usepackage[justification=centering]{caption}

\begin{document}

\title[Fairness Hazard Analysis for Socio-Technical Processes: A Multiple-Case Study in Bias-sensitive Organisational Settings]{Fairness Hazard Analysis for Socio-Technical Processes: A Multiple-Case Study in Bias-sensitive Organisational Settings}

\author*[1]{\fnm{Giovanna} \sur{Broccia}}\email{giovanna.broccia@isti.cnr.it}

\author[1]{Lucio Lelii}\email{lucio.lelii@isti.cnr.it}

\author[1]{Roberto Cirillo}\email{roberto.cirillo@isti.cnr.it}

\author[2]{Dario Di Nucci}\email{ddinucci@unisa.it}

\author[3]{Samuel Fricker}\email{samuel.fricker@fhnw.ch}

\author[2]{Fabio Palomba}\email{fpalomba@unisa.it}

\author[1]{Giorgio O. Spagnolo} \email{spagnolo@isti.cnr.it}

\author*[4,1]{Alessio Ferrari}\email{alessio.ferrari@isti.cnr.it}



\affil*[1]{\orgdiv{ISTI}, \orgname{CNR}, \orgaddress{
\city{Pisa}, 
\country{Italy}}}
\affil[2]{\orgdiv{University of Salerno},  \orgaddress{
\city{Salerno}, 
\country{Italy}}}
\affil[3]{\orgdiv{FHNW University of Applied Sciences and Arts Northwestern}, \orgaddress{\city{Windisch}, \country{Switzerland}}}
\affil*[4]{\orgdiv{Trinity College Dublin}, \orgaddress{\city{Dublin}, \country{Ireland}}}




\abstract{
\textbf{Context.} Fairness is increasingly recognised as a first-class requirement in socio-technical processes, where interactions among human actors, software systems, and AI technologies may lead to unfair outcomes in decision-making workflows. If left unaddressed, fairness hazards may accumulate and reinforce systemic bias, highlighting the need to engineer fairness proactively.  
\textbf{Problem.} Despite growing interest in fairness-aware systems, systematic methods for identifying fairness hazards in socio-technical processes and deriving requirements-level mitigations remain limited.
\textbf{Method.} To support fairness-by-design during requirements engineering (RE), Fairness Hazard Analysis (FHA) is introduced as a methodology for systematically identifying, analysing, and mitigating fairness hazards. FHA is first assessed through a proof-of-concept validation conducted via two focus groups. Then, a qualitative multiple-case study involving two organisations examines its applicability in real-world settings. 
\textbf{Results.} The proof-of-concept validation highlighted the benefits derived from the structured nature of the method, and suggested the need to include iterative, dialogic reflection with domain experts. In the multiple case-study where FHA was applied,   
the practitioners involved were positively impressed by the results and confirmed the relevance of the identified fairness hazards (spanning up to 27\% of the process elements), as well as the appropriateness of most of the proposed mitigations, while noting that contextual factors might hinder their implementation. The evaluation also highlighted mitigation patterns, such as independent review and collective decision-making, which can be transferred to different organisations. 
\textbf{Conclusion.} This paper contributes a structured and empirically validated methodology for integrating fairness considerations in RE and preventing systemic bias in socio-technical processes.

}




\maketitle

\section{Introduction}
Fairness is increasingly recognised as a fundamental quality concern in socio-technical processes, where decisions emerge from interactions among human actors, organisational procedures, software systems, and, increasingly, artificial intelligence (AI) technologies. Although AI models in general, and large language models (LLMs) in particular, are often regarded as the primary source of unfairness~\cite{wei2025addressing}, unfair treatment may also arise from organisational practices, human decision-making, and the interplay among these elements~\cite{selbst2019fairness}. If left unaddressed, such fairness issues may propagate throughout the process, leading to systemic bias over time~\cite{glickman2025human}.

Despite its relevance, fairness is often treated as a post-hoc evaluation concern rather than a requirement to be engineered from the outset~\cite{farahani2021adaptive}. Ensuring fairness, however, requires systematic attention comparable to safety and security, calling for requirements engineering (RE) methods to identify and mitigate fairness risks early in the design process.

While research on algorithmic fairness has produced a wide range of metrics and mitigation techniques, most are typically applied at the AI model or dataset level. Consequently, there remains a lack of operational methods that requirements engineers can apply to analyse and address fairness in socio-technical processes~\cite{soremekun2022software}. Recent work has introduced the notion of \textit{fairness debt}, conceptualising fairness issues as liabilities that accumulate when unaddressed and become increasingly difficult and costly to resolve~\cite{de2025software}.  This perspective further highlights the need for methods that enable fairness concerns to be identified early, analysed in terms of how they may propagate across socio-technical processes, and systematically mitigated before they accumulate into systemic bias.

To meet this need, Broccia et al. propose Fairness Hazard Analysis (FHA)~\cite{broccia2026fairness}, an adaptation of hazard analysis methods from safety engineering~\cite{ericson2015hazard}. This analogy is motivated by the conceptual parallel between safety and fairness: just as safety engineering aims to identify and control conditions that could lead to harm, fairness engineering can systematically anticipate and mitigate conditions that may cause inequitable outcomes. By treating fairness issues as hazard-like states that may emerge and propagate across socio-technical processes, FHA provides a structured approach to identifying and mitigating fairness hazards, thereby helping prevent the accumulation of fairness debt~\cite{de2025software}. From a procedural standpoint, FHA first models the process under analysis, its actors, activities, and information flows; it then supports analysts in identifying fairness hazards, examining their consequences, propagation, impact, and likelihood, and defining requirements-level mitigations that modify or introduce controls within the process. The analysis is iterative, allowing hazards and mitigations to be revisited as the process or its context evolves. 


This article builds upon and extends the work by Broccia et al.~\cite{broccia2026fairness}. While Broccia et al. introduced FHA and provided proof-of-concept validation at Technology Readiness Level (TRL)~3, its applicability in relevant organisational environments had not yet been examined. To address this limitation, the present work refines FHA for application to real organisational workflows and evaluates it through a qualitative multiple-case study, thereby advancing its validation to TRL~5\footnote{TRL~4 was not considered a distinct intermediate stage because FHA is an organisational methodology rather than a technological component amenable to laboratory testing. Following its proof-of-concept validation through expert feedback in focus groups, the next meaningful validation step was therefore its application in relevant organisational environments, consistent with TRL~5.}.

FHA has been applied to recruitment processes for both illustration and evaluation purposes. Recruitment provides a representative socio-technical setting because it involves interactions among organisational procedures, human judgement, software tools, and, in some cases, AI-supported activities. It also includes multiple decision points with potentially significant consequences for individuals and groups, making it particularly suitable for analysing how fairness hazards may emerge, propagate, and be mitigated.

The proof-of-concept evaluation used a constructed but realistic exemplar of an AI-assisted recruitment process to support a structured and accessible discussion of FHA with multidisciplinary experts. The multiple-case study instead applied FHA to two real organisational recruitment processes reconstructed from practitioner interviews and validated by organisational personnel. The evaluations therefore differed in their degree of contextual realism, progressing from a plausibly constructed process to workflows embedded in real organisational settings.

The multiple-case study involved two organisations, anonymised in this paper as Organisation~A and Organisation~B. Organisation~A is a private-sector company operating internationally in the development of machinery and automation systems for industrial applications, whereas Organisation~B is a public higher-education institution. The two cases provide complementary organisational settings in which recruitment processes differ in their structure, governance, decision-making practices, and use of supporting technologies. This variation enables FHA to be examined across distinct organisational contexts.

Compared to Broccia et al.~\cite{broccia2026fairness}, the present paper offers the following extensions and associated contributions:
\begin{itemize}
    \item it broadens and refines FHA for application to real organisational socio-technical processes, by incorporating the elicitation and validation of organisational workflows and by clarifying the complementary roles of organisational members and FHA experts in identifying, assessing, and mitigating fairness hazards;
    \item it presents a qualitative multiple-case study involving two real organisations, in which organisational members contribute domain and process knowledge, while FHA experts support the identification, assessment, and mitigation of fairness hazards;
    \item it characterises the types of fairness hazards that emerge when FHA is applied to real organisational workflows, highlighting both recurring patterns across the two cases and hazards that depend on the specific organisational context;
    \item it investigates practitioners' perceptions of the usefulness, clarity, and applicability of FHA for analysing fairness risks, together with the perceived realism of the identified hazards and the feasibility of the proposed mitigations;
    \item it derives a set of lessons learned from the multiple-case study analysis concerning the conditions that facilitate or hinder the application of FHA in organisational settings. These lessons further refine the methodological guidance, identify practical considerations for its adoption, and inform future applications of FHA to other socio-technical processes;
    \item overall, it advances the empirical validation of FHA from proof of concept at TRL~3 to validation in a relevant organisational environment at TRL~5.
\end{itemize}

The remainder of this paper is structured as follows. Sections~\ref{sec:background} and~\ref{sec:related} present the background and discuss related work, respectively. Section~\ref{sec:FHAmethod} describes the FHA methodology in detail. Section~\ref{sec:pofEval} reports the proof-of-concept evaluation, while Section~\ref{sec:FHAimplement} explains how FHA was operationalised for the multiple-case study. Sections~\ref{sec:studyDesign} and~\ref{sec:results} present the study design and results, respectively. Section~\ref{sec:discussion} discusses the implications of the findings, Section~\ref{sec:lessons} presents the lessons learned, and Section~\ref{sec:threats} addresses threats to validity. Finally, Section~\ref{sec:conclusion} concludes the paper and outlines directions for future work.

\smallskip
\noindent\textbf{Replication Package.} We make our replication package available at \cite{zenodoRepl}.

\section{Background}\label{sec:background}
\subsection{Bias, Fairness, and Hazards}

Bias refers to systematic tendencies in data, processes, or decisions that may produce distorted or unequal outcomes. Such tendencies may arise from unequal representation, historical patterns, organisational practices, or cognitive prejudices affecting human judgement~\cite{varona2022discrimination}. In socio-technical processes, bias may therefore originate from human decision-makers, institutional rules, software-supported activities, or the interactions among these components. Not every form of differentiation is necessarily unfair: whether a distinction is acceptable depends on the specific context, the objectives of the process, applicable ethical and legal principles, and its effects on the stakeholders involved~\cite{selbst2019fairness}.
Fairness concerns the equitable treatment of individuals and groups throughout the socio-technical process, rather than merely the absence of bias within an isolated technical component. It must consequently be assessed in relation to the broader organisational and social context in which decisions are made and implemented. 

Although much existing research focuses on fairness in AI systems, where biased outcomes may arise from unrepresentative training data, model assumptions, or evaluation procedures~\cite{gichoya2023ai,pagano2023bias}, AI is only one possible source of unfairness. Organisational procedures, subjective evaluation criteria, unequal access to information, and inconsistent human decisions may also disadvantage particular individuals or groups.

The introduction of AI may nevertheless amplify existing fairness concerns or create new interactions through which bias propagates. For example, AI outputs may influence subsequent human judgements, while human responses may in turn affect the data and decisions produced in later stages, creating feedback loops that reinforce existing disparities~\cite{glickman2025human}.
These feedback effects represent significant hazards: risks that extend beyond technical malfunction to encompass psychological, social, and ethical consequences \cite{chen2023ethical}. When biased AI systems influence human perception and judgment, they may not only distort individual decision-making but also reinforce societal disparities, making the identification and correction of these feedback-driven hazards a critical challenge for responsible design of AI systems \cite{afreen2025systematic}.

\subsection{Fairness Debt} 
Fairness debt was introduced to explain how fairness issues in software systems accumulate when they are not explicitly managed throughout the software lifecycle~\cite{de2025software}. Aligned with the definitions of technical debt~\cite{alves2016identification} and social debt~\cite{tamburri2015social}, fairness debt refers to the latent socio-technical liabilities arising from fairness oversights, omissions, or trade-offs during development and operation.

De Souza Santos et al.~\cite{de2025software} identify several root causes of fairness debt across the software lifecycle: 
(i) cognitive bias, arising from developers’ subjective assumptions; 
(ii) requirements bias, from incomplete or non-inclusive elicitation; 
(iii) design bias, introduced through architectural or interface choices; 
(iv) historical bias, stemming from legacy data that reproduces inequities; 
(v) training bias, due to unrepresentative datasets; 
(vi) model bias, produced by algorithmic simplifications or parameter settings; 
(vii) testing bias, when validation overlooks fairness metrics; and 
(viii) societal bias, reflecting broader structural inequalities in the system’s context.

These causes are not isolated but interdependent, meaning that fairness issues can propagate across lifecycle stages: for example, an unaddressed requirements bias may evolve into design or testing bias downstream. Over time, the accumulation of such debts increases the risk of systemic inequities, reputational damage, and regulatory non-compliance. Although defined in the context of software development, several of these root causes --- particularly cognitive, societal, and requirements bias --- can also emerge within the human components of socio-technical systems. Human decision-makers interacting with software systems may, for instance, over-rely on algorithmic recommendations, apply subjective evaluation criteria, or reproduce social stereotypes. This socio-technical interpretation reinforces that fairness debt is not purely a software engineering concern but a property of the entire human–software ecosystem. Hence, it should be treated as a managed and traceable property of socio-technical systems, requiring continuous attention rather than post-hoc correction. 

This work builds on this concept by using the identified root causes of fairness debt as analytical prompts for reasoning about potential fairness hazards. However, we acknowledge that these prompts are not sufficient on their own, as identifying fairness hazards also requires detailed knowledge of the socio-technical process under analysis, provided by organisational members and analysts with complementary perspectives.

\subsection{Hazard Analysis}
Hazard analysis is a foundational concept in system safety engineering, aimed at identifying and mitigating conditions that could lead to undesired or unsafe system states~\cite{ericson2015hazard}. A hazard is typically defined as a state or set of conditions that, together with certain triggers, can result in harm or loss~\cite{leveson1995safeware}. The purpose of hazard analysis is to anticipate such conditions as early as possible, evaluate their causes and potential consequences, and design appropriate preventive or corrective controls~\cite{lutz1993analyzing}.

Among the most commonly used hazard analysis techniques, the Preliminary Hazard Analysis (PHA) is a qualitative, top-down approach that provides an initial overview of potential hazards, even before detailed system design information is available \cite{ericson2015hazard}. Its objective is to capture early insights concerning potential risk sources, their likely causes and effects, and to propose preliminary mitigation strategies, which are documented in a hazard table. A PHA generally follows a structured sequence of activities. The process begins with system definition, followed by the identification of potential hazardous conditions, failures, and actions. Each identified hazard is then examined to determine its possible causes and the severity and likelihood of its potential consequences. The combination of severity and likelihood provides a preliminary basis for assessing risk and prioritising hazards that require further attention. Finally, preventive or control measures are proposed to eliminate each hazard or mitigate its associated risk to an acceptable level. The process is iterative: as the system design matures, the identified hazards and the proposed mitigations can be refined.

Building on PHA, Fairness Hazard Analysis (FHA) adapts its qualitative, top-down structure to analyse fairness in socio-technical processes. FHA treats fairness deficiencies as hazard-like conditions, retaining the PHA structure to identify, trace, and mitigate fairness hazards across socio-technical processes.

\section{Related Work}\label{sec:related}

Recent studies have begun to explore how fairness and broader human values can be operationalised throughout the system lifecycle. \emph{Values@Runtime} proposes mechanisms to capture and adapt to stakeholder values during operation~\cite{bennaceur2023valuesruntime}, while \emph{ReFair} focuses on fairness-requirement elicitation in machine learning systems through a context-aware recommender system~\cite{ferrara2024refair}. Empirical analyses further show that fairness is still treated as a secondary quality attribute: developers lack systematic, lifecycle-oriented methods to specify, trace, and maintain fairness requirements~\cite{palomba2024fairaware,voria2025survey}. 
In parallel, safety and security engineering provide well-established hazard analysis frameworks for the early identification and mitigation of risks~\cite{leveson2011engineering}. Recent work demonstrates that system-safety methods can also uncover social and ethical risks in machine-learning systems~\cite{rismani2023stpa}. However, explicit translations of fairness risks into actionable, requirements-level controls across human–AI workflows remain scarce.

Goal-oriented requirements engineering approaches such as i* \cite{yu1995modelling} and KAOS~\cite{dardenne1993goal} provide powerful abstractions for modelling stakeholder goals, dependencies, and obstacles, and have been successfully used to reason about non-functional concerns through softgoals and constraints. However, these approaches primarily focus on goal satisfaction and conflict resolution, and provide limited support for analysing how risks, e.g., fairness issues, may emerge, propagate, and accumulate across socio-technical workflows over time. FHA is therefore positioned as complementary: while goal-oriented models capture what the system should achieve, FHA focuses on identifying and controlling fairness-related hazard conditions arising from interactions between human and AI actors.

FHA systematically identifies fairness hazards by combining multiple sources of evidence and expertise, including empirically established sources of fairness debt, organisational policies and practices, regulatory and ethical guidance, domain-specific knowledge, and the perspectives of stakeholders involved in the socio-technical process. It then analyses how the identified hazards may propagate across the workflow and derives requirements-level mitigations that are explicitly traceable to the unfair outcomes they are intended to prevent. While FHA is structurally inspired by safety hazard analysis, it goes beyond a direct substitution of ``safety'' with ``fairness'' by treating fairness as a contextual and socio-technical concern, explicitly distinguishing undesirable bias from contextually justified differentiation, and defining mitigations as changes to organisational procedures, human decision-making activities, software-supported steps, or their interactions, instead of purely technical controls.

\section{Fairness Hazard Analysis Methodology}\label{sec:FHAmethod}
\begin{wrapfigure}{r}{0.55\textwidth}   
\centering
    \includegraphics[width=1\linewidth]{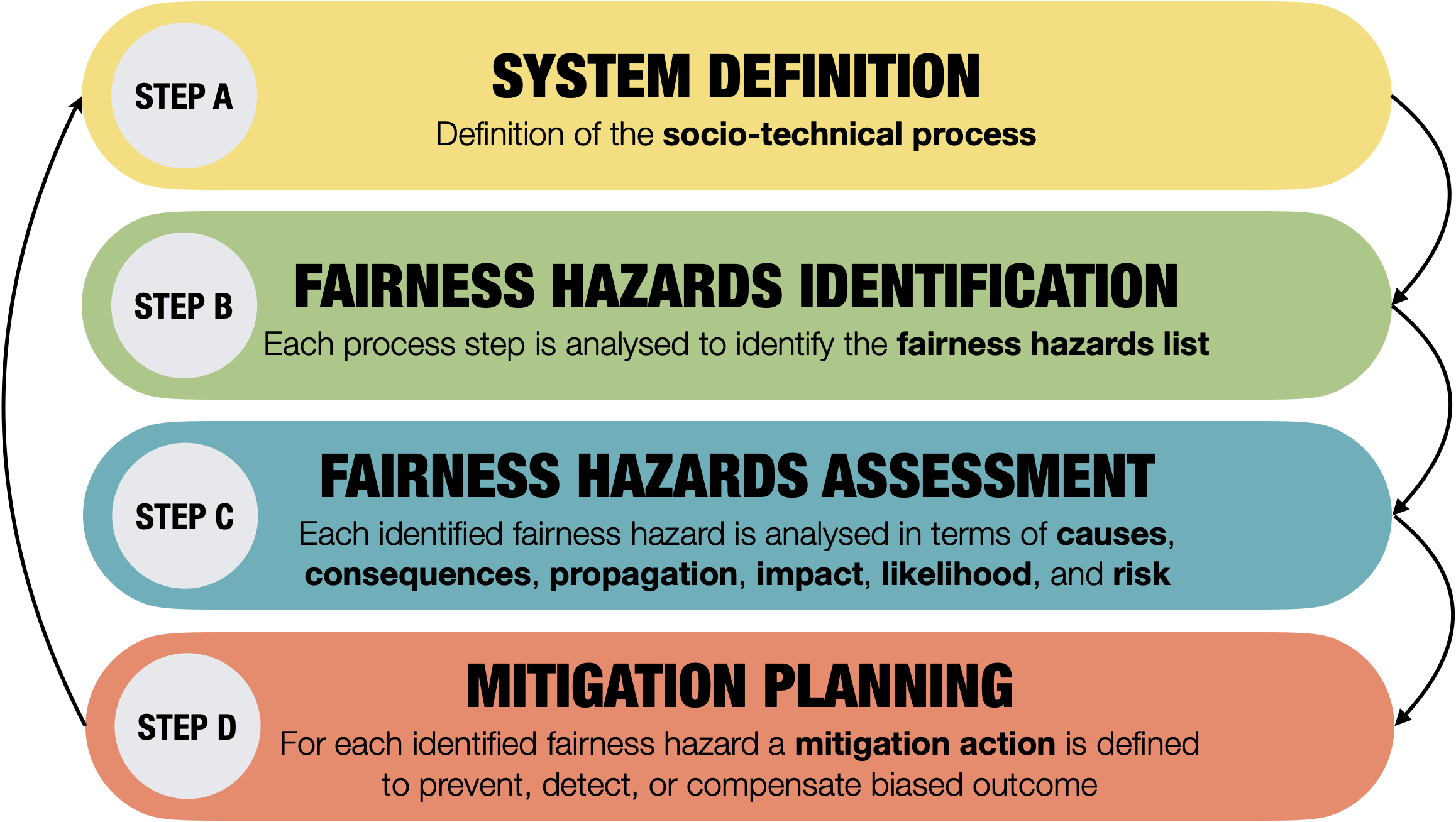}
    \caption{FHA process}
    \label{fig:FHA_diagram}
\end{wrapfigure}

FHA treats fairness issues---e.g., biased decisions, unbalanced access to information, or unequal treatment of agents---as hazard-like conditions that can arise during socio-technical processes, enabling their structured analysis and mitigation at the requirements level. 
Consistent with the structure of PHA, FHA follows the sequential process described below (cfr. Figure~\ref{fig:FHA_diagram}). 
Each step is carried out collaboratively by organisational members with detailed knowledge of the socio-technical process under analysis and by experts in bias and fairness analysis, supported, where relevant, by experts in other pertinent domains, e.g., RE, software engineering, data science, ethics. Their complementary perspectives support the identification of fairness issues, the analysis of how these issues may propagate across the socio-technical process, and the collaborative assessment and mitigation of the resulting hazards.

\smallskip
\textbf{\textit{Step A. System Definition.}} As in PHA, the first step defines and models the socio-technical process, its actors, and their interactions. 

\smallskip
\textbf{\textit{Step B. Fairness Hazard Identification.} }
Each relevant process step, actor, decision point, and interaction is examined to determine whether and how fairness hazards may arise.     
Analysts draw on multiple sources when identifying fairness hazards, including prior empirical studies on fairness debt, organisational knowledge, domain-specific evidence, applicable policies and regulations, the perspectives of the stakeholders involved, and expertise in fairness and bias. FHA supports collaborative sense-making among analysts with complementary backgrounds, allowing hazards to be iteratively refined rather than exhaustively predefined.
This step produces a fairness-hazard list that documents potential fairness issues and their locations.

\smallskip
\textbf{\textit{Step C. Fairness Hazard Assessment.}}
Each identified fairness hazard is analysed in terms of its consequences, propagation, impact, likelihood, and risk through a collaborative review process where analysts discuss and resolve differing judgments to reach consensus.

On the one hand, impact captures the severity of the fairness consequences that may arise if a hazard materialises, ranging from the following values:
\begin{itemize}
\item None: The differentiation is intentional, justified, and does not raise fairness concerns.
\item Low: The hazard may lead to minor unfairness affecting a limited number of individuals or decisions, with negligible consequences on process outcomes.
\item Moderate: The hazard may disadvantage some individuals or groups and influence one or more process decisions.
\item High: The hazard may substantially affect the treatment, opportunities, or outcomes experienced by individuals or groups, resulting in significant unfairness.
\item Critical: The hazard may produce structural or systemic unfairness, potentially violating ethical principles, organisational policies, legal requirements, or societal expectations.
\end{itemize}

On the other hand, likelihood captures the expected frequency with which a fairness issue may occur within the analysed process, ranging from the following values:
\begin{itemize}
\item Rare: The hazard is unlikely to occur and would require exceptional circumstances.
\item Possible: The hazard may occur occasionally under specific conditions.
\item Likely: The hazard is expected to occur regularly because it is associated with recurring activities or decisions within the process.
\item Systemic: The hazard is embedded in the structure, rules, or operations of the process and is therefore expected to recur unless corrective actions are implemented.
\end{itemize}

Combining these aspects allows analysts to prioritise fairness risks (cfr. Table \ref{tab:fairness-matrix}).

\begin{table}[t]
\centering
\caption{Fairness risk matrix obtained by combining impact and likelihood levels.}
\label{tab:fairness-matrix}
\footnotesize
\begin{tabular}{llcccc}
\toprule
& & \multicolumn{4}{c}{\textbf{\textit{Likelihood}}} \\
& & \textbf{Rare} & \textbf{Possible} & \textbf{Likely} & \textbf{Systemic} \\
\midrule

\multirow{5}{*}{\rotatebox[origin=c]{90}{\textbf{\textit{Impact}}}}
& \textbf{None}
& \textbf{None}
& \textbf{None}
& \textbf{None}
& \textbf{None} \\

& \textbf{Low}
& Low
& Low
& Medium
& Medium \\

& \textbf{Moderate}
& Low
& Medium
& Medium--High
& High \\

& \textbf{High}
& Medium
& Medium--High
& High
& Critical \\

& \textbf{Critical}
& Medium
& High
& High
& Critical \\

\bottomrule
\end{tabular}
\end{table}

FHA distinguishes between undesirable bias, which causes harm and requires mitigation, and justified or goal-aligned differentiation, which may be acceptable in context. The latter is recorded for transparency and traceability. While such cases are assigned no risk and require no mitigation, their explicit documentation supports accountability and prevents unexamined assumptions from becoming implicit sources of fairness debt.

This step produces a table reporting each hazard, its consequences, propagation, and qualitative risk classification.

\smallskip
\textbf{\textit{Step D. Mitigation Planning.}} For each fairness hazard, FHA defines control actions at the requirements level that modify the socio-technical workflow to prevent, detect, or compensate for unfair outcomes. Mitigation is achieved by introducing or adjusting workflow nodes and by implementing specific controls at critical decision points. These controls may include procedural additions--such as inserting human review or consensus nodes for high-impact decisions--as well as technical interventions, for instance, refining or constraining AI behaviour through targeted prompt engineering, introducing fairness-aware scoring functions, or enforcing transparency and auditability checkpoints. 
This step produces a list of mitigation strategies, each explicitly linked to the fairness hazard it is intended to address.

\smallskip
As in traditional PHA, FHA is an iterative process. Once the socio-technical process evolves or new empirical evidence emerges, fairness hazards and mitigations are revisited.

\section{FHA Proof-of-Concept Evaluation}\label{sec:pofEval}
To demonstrate FHA and assess the methodology at the proof-of-concept level, Broccia et al. applied it to a constructed, simplified yet realistic exemplar designed to capture key socio-technical characteristics of a recruitment process~\cite{broccia2026fairness}. The process provides a common and controlled basis for explaining the application of FHA, while the evaluation examines the methodology's clarity, perceived usefulness, and areas for improvement. The later multiple-case study, presented in Section~\ref{sec:studyDesign}, complements this evaluation by examining FHA in real organisational settings.

\subsection{Constructed Recruitment Exemplar and FHA Application}
The exemplar represents a simplified AI-assisted recruitment workflow. AI-assisted hiring systems can enhance recruitment quality by improving efficiency and reducing the transactional workload of human personnel. Nevertheless, insufficiently investigated biases may lead to unfair practices and discriminatory outcomes based on factors such as gender, race, ethnicity, or personality traits~\cite{chen2023ethics}.
An AI-assisted hiring process was selected as a representative case because it involves complex, continuous human-AI interactions and decision-making steps that are particularly sensitive to fairness concerns, as demonstrated by previous literature on fairness engineering \cite{fabris2022algorithmic}. Below, we briefly summarise the exemplar and summarise the results of FHA. More details are provided in~\cite{broccia2026fairness}.

\smallskip
\textit{\textbf{Process.}} The process comprises four main components. A \emph{Data Ingestor} receives job descriptions and candidates' curricula and extracts structured requirements and candidate features. An \emph{AI Prescreener} uses these data to rank candidates according to their correspondence with the job requirements. A \emph{Human Recruiter} reviews the original applications and the AI-generated ranking and may confirm or override the recommendations when selecting candidates for interview. Finally, an \emph{Audit Node} records the intermediate artefacts and decisions produced throughout the workflow.

\smallskip
\textit{\textbf{FHA Application.}}  Three analysts applied FHA to the illustrative workflow. The analysis identified eight fairness hazards arising from both human and technical components, including the reproduction of historical inequalities in curricula and job descriptions, incomplete representation of non-standard candidate profiles, an imbalance in the data used by the prescreening system, human overreliance on AI recommendations, subjective or stereotypical human judgements, and others. 
The analysts examined the potential consequences and propagation of each hazard, assessed its impact and likelihood, and defined corresponding mitigation strategies. The proposed controls included validating candidate information, checking job descriptions for exclusionary criteria, documenting potentially relevant historical inequalities, improving the transparency of automated rankings, and introducing independent human reviews and discussion mechanisms for consequential decisions. The purpose of the case was to demonstrate how FHA supports traceability from process steps to hazards, potential unfair outcomes, and requirements-level mitigations; it was not intended to represent a complete or universally applicable recruitment process. Further details of the illustrative analysis are reported in~\cite{broccia2026fairness}.

\subsection{Proof-of-Concept Evaluation}

The proof-of-concept evaluation aimed to investigate whether FHA was understandable and perceived as useful by experts with different disciplinary backgrounds, and to identify aspects requiring refinement before its application in real organisational settings. Given the formative purpose of the evaluation, qualitative focus groups were selected to encourage participants to discuss, challenge, and collectively elaborate on the methodology and its application~\cite{tremblay2010focus}.

\smallskip
\textbf{\textit{Participants.}} The evaluation involved twelve academic participants divided into two focus groups of six participants each. Of the participants, 33.4\% were women and 66.7\% were men. Their prior knowledge of fairness varied: 58\% reported no or basic expertise, while 42\% reported intermediate to advanced expertise. This variation enabled the evaluation to capture the perspectives of both participants already familiar with fairness-related concepts and participants encountering a structured fairness-analysis methodology for the first time.

\smallskip
\textbf{\textit{Procedure.}} Each focus group lasted approximately two hours and was moderated by two researchers. The sessions began with a presentation of the motivation, concepts, and sequential steps of FHA, followed by a walkthrough of its application to the illustrative recruitment case. Participants were encouraged to ask questions and provide observations throughout the presentation.

The original sessions also included a demonstration of a supporting tool under development, HumAInFlow, a no-code, agentic platform designed to model, simulate, and analyse socio-technical workflows in which humans and software, including AI agents, co-exist and collaborate~\cite{HumAInFlow2025}. However, as the present article focuses exclusively on FHA as a methodology, the evaluation reported here considers only the discussion, questions, and findings concerning FHA.  Further details of the tool-specific questions, observations, and recommendations are reported in~\cite{broccia2026fairness}.

Following the presentation, participants discussed a set of open-ended questions addressing the comprehensibility and clarity of FHA, its perceived usefulness for analysing fairness in socio-technical processes, the adequacy of its hazard identification and risk assessment steps, and possible methodological improvements. The questions were intended to stimulate reflection rather than produce quantitative measurements. The complete evaluation instrument (i.e., the slides used with questions) is included in the supplementary material \cite{anonymous_2026_21410249}.
The focus groups were recorded and automatically transcribed. Then, a thematic analysis \cite{braun2006using} was conducted by the last author and was revised by the first author, to identify strengths and points of improvement. 

\textit{\textbf{Results.}} 
Participants generally perceived FHA as a structured and understandable approach for reasoning about fairness. In particular, they valued its ability to direct attention towards fairness issues arising throughout a socio-technical process, rather than restricting the analysis to an AI model or another isolated technical component. One participant observed that `\textit{`the approach feels structured and clear, especially for those familiar with requirements engineering''} and that \textit{``it helps identify fairness issues throughout the process...not only in the AI component.''} 
Table \ref{tab:strengthsFHA} summarises the main strengths of FHA identified during the focus groups, based on the thematic analysis.

\begin{table}[h!]
\centering
\caption{Strengths of the FHA method identified during the focus groups}
\label{tab:strengthsFHA}
\tiny
\begin{tabular}{p{2cm} p{2cm} p{4.2cm} p{3.5cm}}
\toprule
\textbf{Strength Area} & \textbf{Appreciated Characteristic} & \textbf{Description} & \textbf{Exemplary Quote (Participant)} \\
\midrule

\textbf{1. Structured and Systematic Analysis}
& Clear and systematic methodology
& Participants appreciated that FHA provides a structured process for identifying, analysing, and mitigating fairness hazards, making the analysis more comprehensive and repeatable.
& ``The approach feels structured and clear, especially for those familiar with requirements engineering.''\\

& Stepwise hazard analysis
& The staged workflow (hazard identification, assessment, and mitigation) was perceived as intuitive and aligned with established engineering practices.
& ``It is a systematic way to reason about fairness, similar to how we already reason about safety.'' \\

\midrule

\textbf{2. Holistic socio-technical framing}
& Holistic analysis of systems
& Participants valued that the method considers fairness as an emergent property of the entire socio-technical system, rather than focusing exclusively on AI models.
& ``Fairness hazards can emerge not only from a single node, but from the interaction.''  \\

& Integration of human and AI actors
& Explicitly modelling both human and AI decision makers was considered an important strength for understanding where fairness risks originate.
& ``Fairness hazards can emerge from both human and machine actors — this is captured well.'' \\

\midrule

\textbf{3. Awareness and Reflection}
& Promotes fairness awareness
& FHA encourages analysts to explicitly reason about fairness assumptions, biases, and ethical decisions that might otherwise remain implicit.
& ``It makes you aware of biases that otherwise remain implicit.'' \\

& Supports discussion among stakeholders
& Participants appreciated that the method provides a common framework for interdisciplinary discussions about fairness and ethical trade-offs.
& ``Every system should state openly which ethical principles it follows.'' \\

\midrule

\textbf{4. Practical and usable}
& Supports identification of hazards and mitigation strategies
& FHA was appreciated for encouraging systematic reasoning about possible  hazards and mitigation actions and alternative system designs.
& ``The conceptual part is easy to grasp — it follows a logic similar to hazard analysis.'' \\

& Usable reasoning structure
& Participants valued that FHA provides a clear reasoning structure that can guide analysts even when the analysis itself is complex and requires contextual judgement.
& ``Even if complex to apply, the reasoning structure is clear and makes sense.'' \\

\midrule

\textbf{5. Applicability and Generalisability}
& Applicable to real-world domains
& Participants considered the method suitable for realistic socio-technical scenarios such as recruitment, with potential applicability to other domains.
& ``It's important to start with frequent, well-known recruitment cases—that's where this model can bring the most value.'' \\

& Adaptable to different contexts
& The overall FHA process was perceived as sufficiently generic to be transferred to different application domains beyond the running example.
& ``I can see how this could be applied in other contexts beyond hiring.'' \\

\bottomrule
\end{tabular}
\end{table}

The thematic analysis identified five main areas for methodological refinement, together with several associated recommendations.
These themes are summarised in Table~\ref{tab:themesFHA} and described below.

\begin{table}[t]
\centering
\caption{Recommendations for improving the Fairness Hazard Analysis method}
\label{tab:themesFHA}
\tiny
\begin{tabular}{p{0.18\textwidth} p{0.23\textwidth} p{0.23\textwidth} p{0.23\textwidth}}
\toprule
\textbf{Improvement Area} & \textbf{Recommendation} & \textbf{Rationale / Description} & \textbf{Exemplary Quote (Participant)} \\
\midrule
\textbf{1. Clarify and Contextualise Fairness Concepts} 
& Define fairness explicitly for each analysis context 
& The concept of fairness is inherently subjective; the method should require explicit ethical framing and domain-specific definitions. 
& ``For me, the word ‘fairness’ itself is tricky. What’s fair depends on perspective — fairness is inherently biased.'' \\
& Distinguish between acceptable and unacceptable bias 
& The tool could allow users to tag certain biases as “intended” or “undesired” to reflect context-dependent ethics. 
& ``There are desired and undesired biases — for instance, preferring candidates from high-ranking universities might be intentional.''  \\
& Include ethical principle templates 
& Offer pre-defined ethical or fairness frameworks (e.g., distributive justice, equal opportunity) to guide consistent analysis. 
& ``Every system should state openly which ethical principles it follows, so users know what definition of fairness applies.''  \\
\midrule
\textbf{2. Strengthen Bias Detection and Representation} 
& Enhance bias-identification support in the tool 
& Add structured prompts, examples, and checklists for detecting common human and algorithmic biases. 
& ``AI systems often perpetuate existing inequalities, like paying men more than women.''  \\
& Model both human and algorithmic biases distinctly 
& The method should clearly separate bias types and provide visualisation of how they interact. 
& ``You should distinguish between human and machine biases — and possibly even combine their strengths to reduce weaknesses.''  \\
& Support exploration of hidden biases 
& Include sensitivity analysis or simulation tools to uncover biases not explicitly known by analysts. 
& ``But how do we detect biases we don’t know about?''  \\
\midrule
\textbf{3. Improve Analytical Rigour and Usability of FHA} 
& Provide clearer guidance on risk evaluation 
& Develop scales or calibration aids for judging likelihood and impact to reduce subjectivity. 
& ``Judging likelihood and impact is subjective — calibration is needed.'' \\
& Offer domain-specific templates or libraries 
& Create FHA templates for common socio-technical domains (e.g., hiring, healthcare) to ease application. 
& ``It’s important to start with frequent, well-known recruitment cases — that’s where this model can bring the most value.''  \\
& Provide interactive tutorials or example analyses 
& Tutorials can make the structured steps of FHA easier to apply and interpret. 
& ``The approach feels structured and clear, especially for those familiar with requirements engineering.'' \\
\midrule
\textbf{4. Enhance Fairness Mitigation and Iteration Support} 
& Integrate mitigation strategy suggestions 
& When a hazard is identified, the tool could suggest potential mitigation actions (e.g., retraining models, adding review nodes). 
& ``If we know a bias exists — for example, gender bias in historical data — we can retrain models or balance datasets to mitigate it.'' \\
& Promote human–AI collaboration mechanisms 
& Explicitly model roles for human oversight, such as review checkpoints or multi-human consensus steps. 
& ``Use two human recruiters and a discussion node to reduce over-reliance on AI.''  \\
& Support documentation of mitigation rationale 
& Encourage users to record why certain actions were chosen, increasing transparency and accountability. 
& ``The tool allows process simulation to uncover unexpected biases through analysis of outputs.'' \\
\midrule
\textbf{5. Support Reflective and Ongoing Fairness Practice} 
& Encourage iterative, dialogic reflection 
& Build feedback mechanisms for revisiting fairness assumptions as systems evolve. 
& ``Fairness itself must be contextually defined.'' \\
& Frame bias as a learning opportunity 
& Treat the discovery of bias as a positive step toward ethical improvement, not merely a flaw. 
& ``We’re all biased about what counts as bias! Some biases might align with ethical values or goals.'' \\
\bottomrule
\end{tabular}
\end{table}

\begin{description}
    \item[Clarifying and contextualising fairness.] Participants emphasised that fairness cannot be interpreted independently of the organisational, social, ethical, and legal context of the process under analysis. They recommended requiring analysts to state explicitly the interpretation of fairness adopted in each application and to distinguish undesirable bias from intentional or contextually justified differentiation. Ethical principles or domain-specific fairness frameworks were suggested as possible resources for supporting this contextualisation.

    \item[Strengthening hazard identification and representation.]
    Participants considered it important to distinguish hazards arising from human judgement, organisational procedures, software systems, and AI technologies, while also representing their interactions. They recommended providing structured prompts, examples, or checklists to help analysts consider different potential sources of bias. At the same time, participants recognised that predefined categories cannot ensure the identification of unknown or strongly context-dependent hazards.
    \item[Improving analytical rigour and applicability.]
    Although participants considered the sequential structure of FHA understandable, they noted that assessing impact and likelihood necessarily involves subjective judgement. They therefore recommended clearer definitions, calibration guidance, and examples to support more consistent risk assessments. Domain-specific examples, templates, and previously documented hazards were also suggested as resources that could facilitate the application of FHA without replacing contextual analysis.
    \item[Supporting mitigation and iteration.]
    Participants valued the explicit connection between identified hazards and mitigation strategies. They recommended supporting analysts with examples of potential controls, while preserving the need to adapt each mitigation to the organisational context. Particular attention was given to human-oversight mechanisms, such as independent reviews, decision checkpoints, and multi-person consensus for high-impact decisions. Participants also recommended documenting the rationale for selecting, rejecting, or adapting each mitigation.
    \item[Promoting reflective and ongoing fairness practice.]
    Participants viewed fairness analysis as an iterative activity rather than a one-off assessment. They recommended revisiting hazards, assumptions, and mitigation strategies as organisational processes, technologies, regulations, or stakeholder expectations evolve. The identification of bias was also framed as an opportunity for organisational reflection and learning rather than solely as evidence of a process failure.
\end{description}

The proof-of-concept evaluation was designed to obtain formative qualitative feedback and is subject to several limitations. All participants were academics. Consequently, the findings do not fully represent the perspectives and constraints of organisational practitioners. Additionally, the evaluation relied on a single constructed recruitment case rather than workflows elicited from real organisations. The perceived clarity and usefulness of FHA might therefore differ when the methodology is applied to more complex, real processes. These findings motivated the validation of FHA real-wrold organisational multiple-case study described in Section~\ref{sec:studyDesign}.
\section{FHA Operationalisation in the Multiple-Case Study}
\label{sec:FHAimplement}
\begin{figure}
    \centering
    \includegraphics[width=0.7\linewidth]{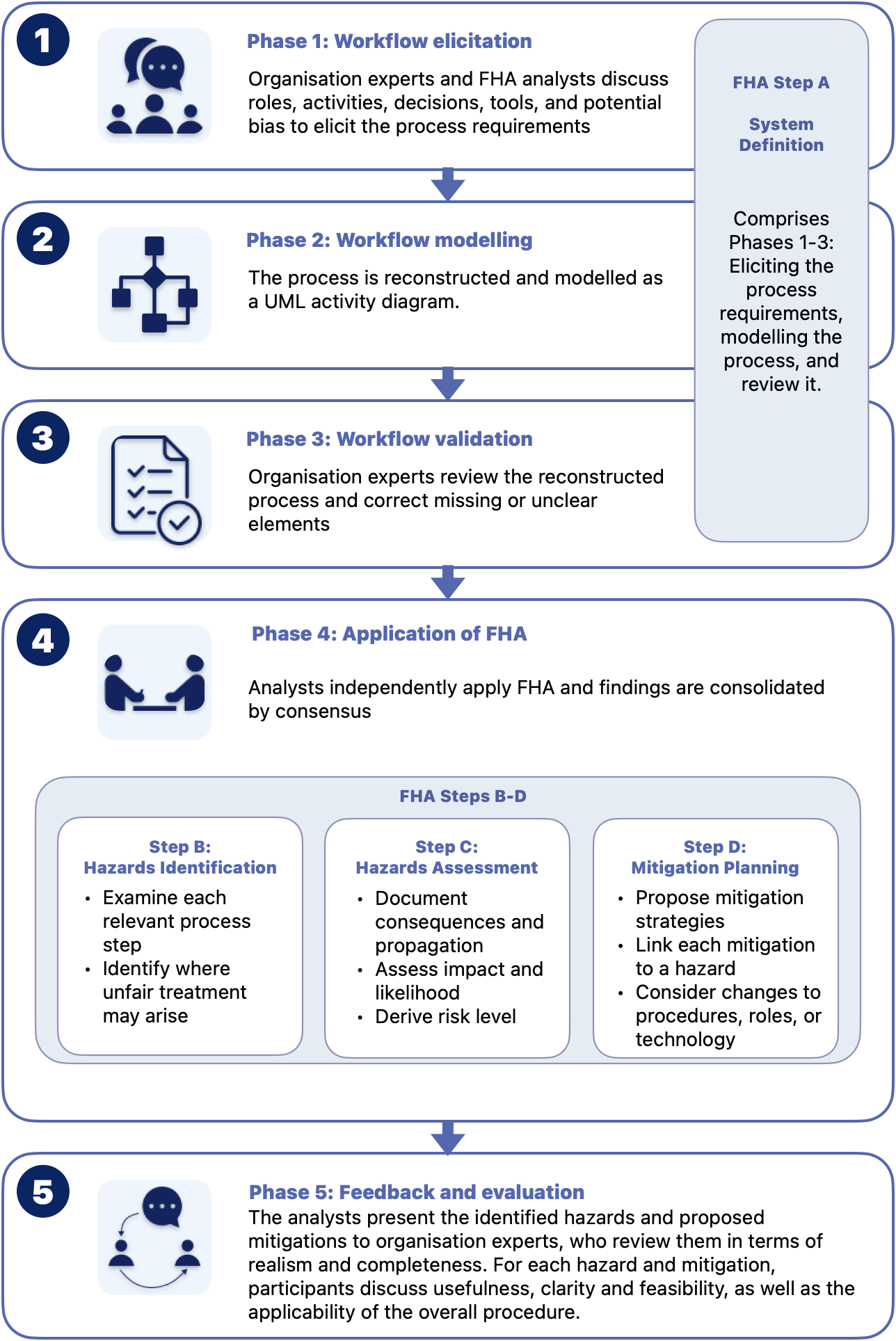}
    \caption{FHA Operationalisation in the Multiple-Case Study}
    \label{fig:FHAimpl}
\end{figure}

To examine FHA in relevant organisational environments, we operationalised the methodology through an analyst-led and practitioner-informed procedure, applied separately to each organisation. Figure \ref{fig:FHAimpl} shows the entire operationalisation.

At this stage of maturity, organisational participants were not expected to apply FHA independently. Instead, they contributed detailed knowledge of the recruitment process and its organisational context, validated the reconstructed workflows, and assessed the resulting fairness hazards and mitigation strategies. The formal application of FHA was conducted by two researchers acting in complementary roles. 
The procedure comprised five sequential phases, applied independently to each organisational case.

\smallskip
\textit{\textbf{Phase 1: Exploratory focus group and workflow elicitation.}}
The first phase aimed to elicit the recruitment process currently adopted by the organisation. Organisational participants were selected for their direct involvement in and detailed knowledge of the process under analysis. Each session was facilitated by one researcher, while a second researcher acted as observer and note-taker. With the participants' consent, the sessions were audio-recorded and subsequently transcribed.
At the beginning of each session, the researchers introduced the study objectives, the notion of a socio-technical process, the concept of a fairness hazard, and the FHA methodology at a high level. The discussion then addressed the main process phases, the organisational roles involved, manual and software-supported activities, decision points, information exchanged and artefacts produced, and the software or AI technologies used. Participants were also asked about process steps they considered particularly sensitive, stakeholders who might be unintentionally disadvantaged, existing organisational safeguards, and the potential influence of human or technological bias. Where AI-supported activities were present or envisaged, the discussion also considered the influence of AI outputs, the required levels of human supervision, transparency needs, and the risks of overreliance.

\smallskip
\textit{\textbf{Phase 2: Workflow reconstruction and modelling.}}
Based on the focus group transcript and collected notes, the analysts reconstructed the recruitment workflow used by each organisation. The resulting processes were modelled as UML activity diagrams. The diagrams represented the activities and decision points of each process, the organisational roles responsible for them, the software and AI technologies involved, and the information, artefacts, and decisions exchanged across process steps. Activity and decision nodes captured the relevant process steps and interactions performed or supported by human actors, software tools, or AI technologies, while control and object flows represented the sequence of activities and the exchange of information, artefacts, and decisions. 

\smallskip
\textit{\textbf{Phase 3: Workflow validation.}}
Before applying FHA, the reconstructed workflow was reviewed by the organisational participants. They were asked to identify inaccuracies, missing steps, unclear responsibilities, and incorrect or incomplete information flows. The analysts incorporated the resulting corrections and clarifications into the process model. FHA was applied only after the participants had confirmed that the representation adequately reflected the organisational process. This phase reduced the risk of fairness hazards being inferred from an inaccurate or incomplete reconstruction of the workflow. The reconstructed and revised workflow (i.e., the output of Phases 1-3) provided the system definition required by Step~A of FHA and the basis for analysing where fairness hazards could emerge and how their consequences might propagate.

\smallskip
\textit{\textbf{Phase 4: Application of FHA.}}
The formal application of FHA was conducted by two researchers acting in complementary roles. The \emph{FHA analyst}, who had led the development of the methodology, and the \emph{supporting analyst}, who was knowledgeable about FHA, independently applied FHA to each validated workflow. Each analyst worked on a printed copy of the corresponding UML activity diagram, using pen-and-paper annotations to mark potential fairness hazards, affected process steps, and preliminary mitigation ideas. After the independent analyses, the
analysts compared the identified hazards, discussed similarities and differences, and jointly reviewed the rationale behind each finding. Hazards identified by both researchers were retained, while those identified by only one researcher were discussed until consensus was reached on their relevance, scope, and formulation. 

This phase operationalised Steps~B--D of the FHA methodology described in Section~\ref{sec:FHAmethod}. 
In \textbf{Step~B}, \emph{Fairness Hazard Identification}, each relevant process step was examined to determine whether it could give rise to unfair treatment. 
In the illustrative exemplar presented by Broccia et al.~\cite{broccia2026fairness}, the recruitment process was intentionally simplified, and each fairness hazard could be directly associated with a workflow node, with each node representing a single actor performing a single, well-defined activity. By contrast, the real organisational recruitment processes analysed in this study included conditional branches, decision gateways, and activities involving multiple actors. Consequently, fairness hazards could not always be naturally associated with an individual workflow node. For the purposes of this operationalisation, we therefore adopted the broader notion of a \emph{process step}, referring to any activity, decision point, or workflow segment in which a fairness hazard may emerge, propagate, or influence subsequent decisions. This adaptation preserves the underlying FHA methodology while supporting its application to more complex socio-technical processes.

Hazard identification drew on the organisational information elicited from participants, the analysts' expertise in fairness and bias, applicable organisational practices and safeguards, the analysts' prior experience of recruitment processes both as participants and evaluators, and the specific characteristics of the recruitment process.

In \textbf{Step~C}, \emph{Fairness Hazard Analysis}, the analysts documented, for each identified hazard, its potential consequences, the downstream decision point or activity that could be directly affected, and the mechanism through which the issue could propagate. Although all identified fairness hazards may ultimately influence the final hiring outcome, the propagation analysis focused on the first downstream decision point directly affected by each hazard. This makes it possible to identify the latest stage at which the process can still be redirected, revised, or corrected before the resulting unfairness propagates further. The analysts then assessed each hazard's impact and likelihood and derived the corresponding qualitative risk level\footnote{In the current study, we do not provide detailed calibration guidelines for impact and likelihood assessment, nor do we evaluate the reliability or soundness of the resulting ratings. Establishing such guidance requires a dedicated study involving sustained participation from experts in fairness, risk assessment, the application domain, and the organisational process. Techniques such as structured expert elicitation, the Delphi method, or nominal-group procedures may be used to iteratively compare judgements, clarify disagreements, and develop shared assessment criteria under uncertainty~\cite{mcmillan2016use}. The limited availability of the required experts made such an evaluation infeasible at the present stage. We therefore defer the systematic calibration and empirical evaluation of Step~C to subsequent validation at TRL~6 and TRL~7.}.

In \textbf{Step~D}, \emph{Mitigation Planning}, the analysts proposed one or more requirements-level mitigation strategies for each fairness hazard. Each mitigation was explicitly linked to the hazard it was intended to address and could involve changes to organisational procedures, decision criteria, human responsibilities, review mechanisms, software-supported activities, or interactions among these elements.

The outputs of this phase comprised the fairness-hazard list, the hazard-analysis and risk-assessment tables, and the corresponding mitigation strategies.

\smallskip
\textit{\textbf{Phase 5: Feedback and evaluation session.}}
The analysis was presented to the organisational participants during a dedicated feedback session. The workflow was reviewed incrementally, one fairness hazard at a time. For each hazard, the researchers presented the relevant activities and actors, the identified fairness hazard potential consequences, and the proposed mitigation strategies.

Participants were invited to assess whether each hazard represented a realistic risk in their organisation, whether any risk appeared exaggerated or implausible, and whether relevant hazards had been overlooked. For each mitigation, they discussed its feasibility, the practical difficulties of implementation, and the organisational barriers that could hinder adoption. After reviewing the complete workflow, participants reflected on the clarity, usefulness, and applicability of FHA, including whether the analysis had surfaced fairness concerns not previously considered and whether it could improve organisational awareness of fairness risks. The sessions were audio-recorded and transcribed with participants' consent for subsequent qualitative analysis.

\section{Multiple-Case Study Design}\label{sec:studyDesign}
The multiple-case study applied the methodology operationalised as specified in the previous section, and it is reported with reference to the guidelines for
case study research in software engineering proposed by Runeson and
Höst~\cite{runeson2009guidelines}. The study follows a holistic multiple-case design\footnote{A holistic case-study design examines each case as a whole, without defining separate embedded units of analysis within it~\cite{runeson2009guidelines}.} comprising two cases, each corresponding to a real recruitment process situated within its organisational context. 

\smallskip
\textbf{\textit{Study Objective and Research Questions.}}
The objective of the multiple-case study is to evaluate the applicability of FHA in relevant organisational environments to address the following research questions:
\begin{description}
\item[\textbf{RQ1.}] What types of fairness hazards emerge when FHA is applied to real organisational recruitment workflows, and what types of mitigation strategies can be proposed to address them?

\item[\textbf{RQ2.}] How do practitioners perceive the usefulness, clarity, and applicability of FHA for analysing fairness risks in socio-technical processes?

\end{description}

To answer the RQs, each of the two recruitment workflows was analysed as a whole through FHA. To address RQ1, the two workflows were first analysed separately to identify and categorise the fairness hazards and corresponding mitigation strategies within each organisational context. The findings from the two cases were then compared through a cross-case synthesis to identify recurring and context-specific categories of hazards and mitigations. 

Organisational personnel served as study subjects by contributing knowledge of the recruitment processes, validating the reconstructed workflows, and evaluating the resulting FHA analyses. To address RQ2, their feedback was first examined within each organisational case and then synthesised across the cases through thematic analysis to identify recurring and context-specific perceptions concerning the usefulness, clarity, and applicability of FHA.

\smallskip
\textbf{\textit{Case and Subjects Selection.}} Two organisational cases were selected by striving for maximum variation, while ensuring that both involved formal hiring processes with multiple actors and decision points. One case organisation was a well-established global technology company headquartered in Belgium, and the other one a young school of a public higher-education institution located in Switzerland. The former organisation was represented by a director responsible for company-wide human resources, and the latter was represented by an experienced human resources specialist team. The choice of these two very different organisations ensured that the case study contexts offered insights in the application of FHA in differing organisational structure, culture, governance, and recruitment procedures. The variation enabled the analysis to examine whether similar categories of fairness hazards and mitigation strategies emerged across distinct organisational contexts.

\smallskip
\textbf{Case A} is a well-established Europe-based international manufacturer of machines, systems, and turnkey solutions, now operating through multiple business regions with a worldwide sales and service network and distribution in more than 50 countries. It is publicly listed at a European stock exchange. Its product portfolio includes cyberphysical systems, finishing equipment, and software platforms. Its 2025 revenue reached more than 500 million Euro and employed approximately 2000 staff. It emphasised teamwork, customer focus, and a safe, collaborative culture with good opportunities for continuous people development.

We applied the FHA method on the recruitment processes of case A organisation by closely collaborating with the company's director of corporate alignment who was responsible for company-wide human resource management. She had more than 16 years of experience in the company and previously had headed several other corporate functions, including head of marketing and head of communications. She brought in-depth knowledge of the company's recruitment procedures and policies. The director participated in workflow elicitation, validation of the reconstructed process, and the feedback session concerning the identified hazards and proposed mitigations.

\smallskip

\textbf{Case B} is a school of a practice-oriented university that combines education, continuing education, and applied research, with a strong focus on digital transformation, AI, data science, IoT, and IT security. It describes itself as inspiring, science-driven, and well-connected, with an emphasis on close industry links and practical impact. The school employs about 150 staff across 3 institutes and 2 campuses. The school was still in a build-and-scale phase: it became operational in 2025 and was continuing to develop structures, programs, and specialist capabilities.

We applied the FHA method on the recruitment processes of case B organisation in close collaboration with the school's lead human resource manager and a human resource expert in her team. The manager had more than 15 years experience in human resources and was with the university for more than three years. The expert had more than 40 years of experience in human resources and finance administration and was for more than 5 years with the university. Together the two participants brought in-depth knowledge and expertise in human resource management and the university's procedures and policies. Both human-resource managers participated in these phases. Their continued involvement supported consistency between the process information initially elicited, the validated workflow, and the subsequent evaluation of the FHA results.

\smallskip
\textbf{\textit{Data Collection.}}
Data collection was conducted separately for each organisational case and followed the operationalisation protocol described in Section~\ref{sec:FHAimplement}. It comprised workflow elicitation, process clarification, FHA application, and practitioner feedback. Table~\ref{tab:case-study-phases} summarises the participants involved, the evidence collected or produced, and the purpose of each phase.

The information about each recruitment process was collected through a two-hour semi-structured elicitation session conducted via videoconference with the organisational participants.
The workflow of Organisation~A was elicited through an individual semi-structured interview, whereas the workflow of Organisation~B was elicited through a semi-structured group interview with the two participating human resource managers.
The discussion covered the main activities and decision points of the process, the organisational roles involved, the information and artefacts exchanged, the software tools and AI technologies used, and any exceptions or alternative process paths. Participants were also asked to clarify responsibilities, decision criteria, existing safeguards, and aspects of the process that could not be represented unambiguously. The sessions were audio-recorded with the participants' consent and transcribed to support workflow representation and subsequent analysis. The collected information was used to define a first UML activity diagram representing the workflow. 

The reconstructed recruitment process was discussed via email with the same organisational participants to clarify and validate the elicited information. The researchers presented a textual description of the process and asked targeted questions concerning points that remained uncertain or ambiguous. A textual representation was used in place of the diagram, to minimise the communication barrier. The participants' responses were used to refine the UML activity diagram. The resulting workflow constituted the system definition required by Step~A of FHA.

After FHA had been applied to each validated workflow, a feedback session lasting 30 minutes (Organisation A) and two hours (Organisation B) was conducted via videoconference with the participants. During these sessions, the identified fairness hazards and proposed mitigation strategies were presented incrementally in relation to the relevant parts of the workflow. For each hazard, participants were asked whether the risk was realistic, unrealistic, or exaggerated and whether any important fairness risks had been overlooked. For each mitigation strategy, they were asked whether it would be feasible within their organisation and which organisational barriers could hinder its adoption. Participants were also asked to reflect on the clarity, usefulness, and practical applicability of FHA and on whether the analysis had helped them recognise fairness concerns that had not previously been considered. The feedback sessions were audio-recorded and transcribed for analysis.

The evidence used to address RQ1 comprises the validated UML activity diagrams, the independent pen-and-paper annotations produced by the two analysts, the consolidated fairness-hazard lists, the hazard-analysis and risk-assessment tables, and the corresponding mitigation strategies. The evidence used to address RQ2 comprises transcripts of the organisational feedback sessions, along with participants' comments on the accuracy of the identified hazards, the feasibility of the proposed mitigations, and the perceived usefulness, clarity, and applicability of FHA.

In Organisation~A, one participant contributed to all three data-collection activities. In Organisation~B, both participating human-resource managers contributed to workflow elicitation, workflow validation, and the final feedback session. Maintaining the same participants across phases supported continuity among the initially elicited process information, the validated workflow, and the evaluation of the FHA analysis. Although we acknowledge that the limited number of participants involved might hinder our ability to faithfully reconstruct the complete workflows, we highlight that: (1) these participants have primary roles in the processes, and have a holistic perspective on the workflows; (2) we do not aim for the perfect reconstruction of the processes, but at understanding the value and applicability of FHA.

\begin{table}[t]
\centering
\caption{Data collection and analysis phases in the multiple-case study.}
\tiny
\label{tab:case-study-phases}
\begin{tabular}{p{0.2\textwidth} p{0.2\textwidth} p{0.2\textwidth} p{0.2\textwidth}}
\toprule
\textbf{Phase} & \textbf{Participants} & \textbf{Evidence produced} & \textbf{Purpose} \\
\midrule
Workflow elicitation
& Organisational personnel and researchers
& Recording, transcript, and notes
& Reconstruct the recruitment process \\

Workflow clarification
& Organisational personnel and researchers
& Clarifications and refined textual process description and UML activity diagram
& Complete and refine Step~A of FHA \\

FHA application
& Two analysts
& Independent annotations and consolidated FHA tables
& Identify and analyse hazards and propose mitigations \\

Feedback evaluation
& Organisational personnel and researchers
& Recording and transcript
& Evaluate the identified hazards, proposed mitigations, and FHA \\

\bottomrule
\end{tabular}%
\end{table}

\smallskip
\textbf{\textit{Data Analysis.}}
The analysis was conducted in two stages. Each organisational case was first analysed separately to preserve the relationship between the findings and the specific recruitment process in which they emerged. The resulting within-case findings were subsequently compared and synthesised across the two cases. 

To address \textbf{RQ1}, we applied FHA to the two validated recruitment workflows, as described in Section~\ref{sec:results}. For each case, the consolidated FHA analysis provided a list of fairness hazards, their potential consequences and propagation, their assessed impact and likelihood, and the corresponding mitigation strategies. Comments provided by organisational participants during the feedback sessions were used to clarify or refine the formulation of hazards and mitigations where necessary.
The final hazards from each case were then categorised by grouping entries that reflected similar sources, mechanisms, or forms of potential unfairness.
For instance, hazards concerning exceptional hiring procedures, internal recruitment paths, or limited publication channels were grouped as instances of \emph{unequal access to opportunities}, because they may give some candidates privileged access to employment opportunities while excluding or reaching others less effectively. Hazards concerning familiarity with candidates or with candidates' background, subjective evaluation constructs, or insufficiently validated assessment criteria were instead
grouped as instances of \emph{subjective or affinity-based evaluation}, since the evaluation may be influenced by
recruiters' prior knowledge, perceived similarity to the candidate, personal
impressions, or criteria that are not consistently defined or applied. 

Mitigation strategies were categorised separately according to the type of intervention proposed, such as changes to organisational procedures, decision criteria, human responsibilities, review mechanisms, or software- and AI-supported activities. 
For instance, mitigations requiring explicit responsibility allocation, periodic oversight, documentation of deviations, or justification for exceptional and consequential decisions were grouped as instances of \emph{governance and accountability}, since they aim to make decision-making responsibilities explicit, support traceability, and ensure accountability for deviations from the standard process. Mitigations requiring standardised artefacts, structured rubrics, predefined evaluation criteria, scoring rules, or documented evaluation rationales were instead grouped under \emph{standardisation and structured evaluation}, since they aim to reduce variability and subjectivity in candidate assessment.

The categorisation was conducted iteratively: entries were compared, candidate categories were created or refined, and conceptually similar entries were grouped under a shared category while preserving relevant differences between them. One researcher performed the initial categorisation, while a second reviewed and double-checked the resulting categories. 

After the within-case categorisation, a cross-case comparison was performed. A comparison matrix was used to examine which categories of fairness hazards and mitigation strategies occurred in both organisational cases and which were specific to one recruitment process. Categories represented in both cases were treated as recurring across the analysed recruitment contexts, whereas those found in only one case were examined in relation to the organisational characteristics and process steps from which they emerged. The cross-case findings were subsequently integrated through a narrative synthesis \cite{cruzes2015case} describing common patterns, contextual differences, and relationships between hazard and mitigation categories.

To address \textbf{RQ2}, we performed thematic analysis \cite{braun2006using} of the transcripts from the organisational feedback sessions. The analysis focused on participants' perceptions of the identified hazards, the proposed mitigation strategies, and FHA as a methodology. The transcripts were first examined separately for each organisational case. Relevant passages were coded according to the issues raised by participants, including the realism and completeness of the identified hazards, the feasibility of the proposed mitigations, potential barriers to their adoption, and the perceived usefulness, clarity, and applicability of FHA.
For example, passages in which participants confirmed that a hazard could occur in their organisation were coded as \emph{hazard realism}, whereas passages in which they qualified a hazard because of existing safeguards, formal procedures, or contextual conditions were coded as \emph{contextual condition} or \emph{existing safeguard}. Similarly, comments concerning whether a proposed mitigation could be implemented were coded as \emph{mitigation feasibility}, while comments referring to cost, candidate burden, scheduling difficulties, or compatibility with existing practices were coded as \emph{adoption barriers}. Passages in which participants stated that the analysis made them reflect on risks or safeguards not previously considered were coded as \emph{awareness raising}.
The codes were iteratively reviewed and grouped into broader themes representing recurring patterns in participants' feedback.
For instance, codes such as \emph{hazard realism}, \emph{contextual condition}, \emph{existing safeguard}, and \emph{hazard completeness} were grouped under the theme \emph{context-dependent validation and completeness of hazards}. 

A cross-case thematic comparison was then conducted to identify perceptions shared across the two organisations and observations that appeared specific to one organisational context. For example, comments from both organisations concerning the feasibility of mitigation strategies were compared to determine whether feasibility concerns reflected a shared issue, such as resource constraints, or a context-specific concern, such as compatibility with a particular recruitment practice. A case-by-theme matrix was used to retain the provenance of the evidence and avoid obscuring differences between the cases during synthesis. 

One researcher conducted the initial coding and thematic synthesis, while a second researcher reviewed and double-checked the coded passages, the resulting themes, and the cross-case interpretation. Representative quotations were selected to illustrate each theme and to maintain a direct connection between the reported interpretations and the participants' statements.

The findings from RQ1 and RQ2 were finally integrated in the Discussion by
relating the hazard and mitigation categories identified through FHA to the
practitioners' feedback on their realism, feasibility, and organisational
relevance. This integrative interpretation provided complementary evidence
concerning the applicability of FHA in real organisational environments.

\smallskip
\textbf{\textit{Ethical Considerations.}}
Before each session, participants were informed about the purpose of the study and asked for permission to record the videoconference for transcription and analysis. To protect confidentiality, participants and organisations were anonymised, and identifying details were removed or generalised in the reported findings and quotations. The research protocol was submitted to FHNW University of Applied Sciences and Arts Northwestern
by following the applicable ethics approval procedures. The protocol had been approved and double-signed before the case study interviews were started.

\section{Results of the Case Study}\label{sec:results}

\subsection{Organisational Recruitment Workflows}

\begin{figure}[h!]
\centering
\begin{minipage}[t]{0.6\textwidth}
    \vspace{0pt}
    \includegraphics[width=\linewidth]{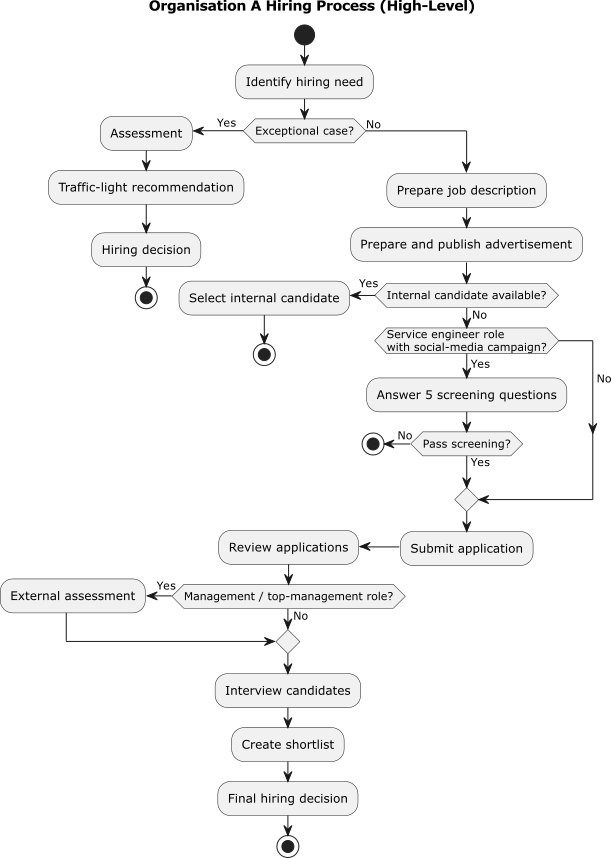}
\end{minipage}
\hfill
\begin{minipage}[t]{0.35\textwidth}
    \small
    \textbf{Process Overview.}
The recruitment process starts when a hiring need is identified, or an exceptional hiring situation arises. In the standard process, an existing job description is reused, or a new one is created and, when requested, refined. Before opening the position externally, the organisation evaluates whether a suitable internal candidate is available. If not, the vacancy is published through the company website and additional recruitment channels. Candidates submit their applications, which are manually reviewed by the personnel manager and/or hiring manager. Selected candidates participate in one or more interview rounds. The process concludes with the creation of a shortlist and a final hiring decision made jointly by the hiring manager and the superior manager.

\textbf{Process Variants.}
The process includes several variants. Service-engineer positions may use a social-media campaign with pre-application screening questions. Management and other key positions may require an external assessment producing a traffic-light recommendation. In exceptional hiring situations, candidates may follow a dedicated assessment path that also results in a traffic-light recommendation before the final hiring decision.
\end{minipage}

\caption{Simplified representation of the Organisation A hiring process.}
\label{fig:hrProcOrgA}
\end{figure}

\begin{figure}[h!]
\centering

\begin{minipage}[t]{0.56\textwidth}
    \vspace{0pt}
    \includegraphics[width=\linewidth]{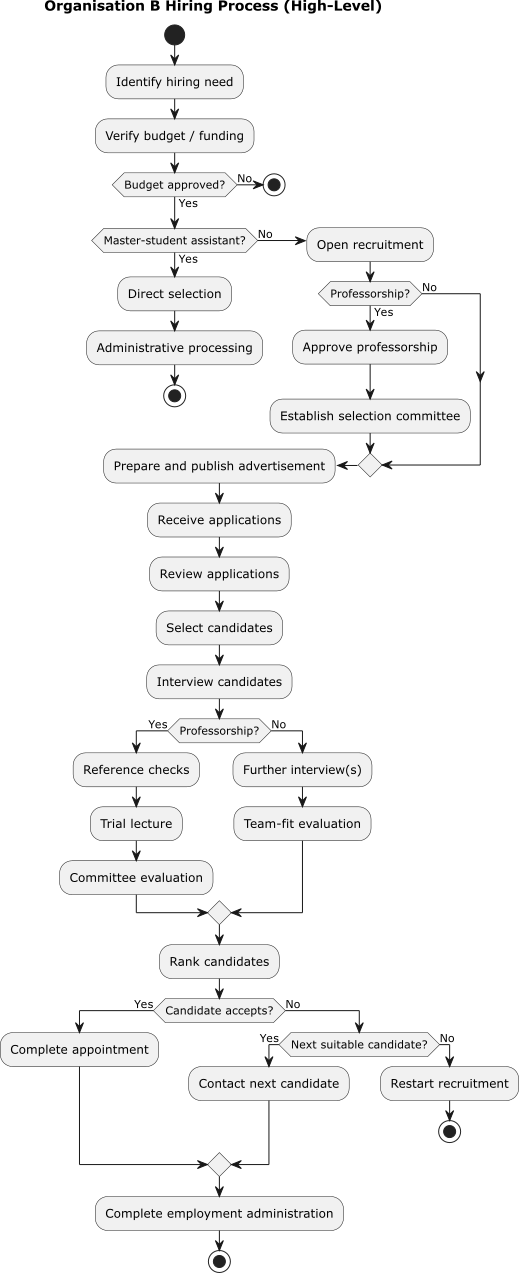}
\end{minipage}
\hfill
\begin{minipage}[t]{0.35\textwidth}
\small
\textbf{Process Overview.}
The recruitment process starts when a hiring need is identified, and budget availability is verified. For standard positions, a recruitment request is submitted, and the vacancy is prepared and published through the RefLine recruitment platform. Candidates submit their applications, which are reviewed by HR and the responsible managers. Selected candidates participate in one or more interview rounds, after which candidates are ranked. If the preferred candidate declines the position, the next suitable candidate may be contacted. The process concludes with the formal appointment and the completion of the employment administration.
\textbf{Process Variants.}
The process includes two main variants. Master-student assistants may follow a simplified recruitment process that involves direct candidate identification and administrative processing, without the standard recruitment workflow. Professorships require additional approval before recruitment begins and involve a formal selection committee responsible for evaluating candidates through dedicated activities, including reference checks, trial lectures, committee discussions, and voting to rank the candidates.
\end{minipage}
\caption{Simplified representation of the Organisation B hiring process.}
\label{fig:hrProcOrgB}
\end{figure}

The outputs of Phases~1--3 of the FHA operationalisation are the reconstructed
and validated organisational workflows. These workflows constitute the system
definition required by Step~A of FHA and provide the basis for identifying and
analysing fairness hazards and for developing mitigation strategies to address
RQ1.
Figure~\ref{fig:hrProcOrgA} presents a simplified version of the Organisation A workflow, together with its description. The complete diagram (comprising 35 activity nodes and 10 decision points distributed across 12 partitions) is available in \cite{anonymous_2026_21410249}. Figure~\ref{fig:hrProcOrgB} presents a simplified version of the Organisation B workflow, while the complete diagram (comprising 58 activity nodes and 12 decision points distributed across 20 distinct organisational role partitions) is available in the replication package \cite{anonymous_2026_21410249}. 

\subsection{RQ1: Fairness Hazards and Mitigation Strategies}
To address RQ1, Steps~B--D of FHA were applied separately to the validated recruitment workflow of each organisation, following the operationalisation described in Section~\ref{sec:FHAimplement}. 
The main text of this section reports the outputs of Steps~B and~D. First, the fairness hazards identified in Step~B are presented alongside the process steps in which they may arise. The corresponding mitigation strategies developed through Step~D are then reported and explicitly linked to the hazards they are intended to address. Step~C remains an essential part of FHA because it supports the assessment of potential consequences, propagation, impact, likelihood, and qualitative risk, thereby enabling analysts to prioritise the hazards requiring mitigation. In the present study, however, mitigation strategies were proposed for all identified hazards rather than only for those assigned higher risk levels. To preserve readability and avoid introducing additional extensive tables, the detailed Step~C analyses are not reproduced in the paper; instead, they are provided in the replication package.

Sections~\ref{sec:rq1OrgA} and~\ref{sec:rq1OrgB} present the within-case results for Organisations~A and~B, respectively. The subsequent cross-case synthesis (see Section~\ref{sec:rq1synthesis}) categorises and compares the identified hazards and mitigation strategies to distinguish recurring patterns from findings specific to one organisational context.

\subsubsection{Organisation A}\label{sec:rq1OrgA}
The hazard-identification procedure (Step B in FHA) for Organisation A resulted in the fifteen fairness hazards (FH-A\textit{i}) reported in Table~\ref{tab:orgA_hazards}. Of these hazards, $7$ were identified independently by both analysts. The remaining $8$ were initially identified by only one analyst: $2$ by one analyst and $6$ by the other. All analyst-specific hazards were discussed during the consolidation phase and retained after consensus was reached on their relevance, scope, and formulation. 
The identified hazards span the entire recruitment process, affecting approximately
27\% of the distinct process elements in the validated workflow. A first category concerns process governance, including the absence of an explicit responsibility for verifying the consistent execution of fairness-relevant activities throughout the recruitment process (FH-A1). Several hazards affect access to employment opportunities, including exceptional hiring situations (FH-A2), exclusionary or tailored job descriptions (FH-A3--FH-A4), AI-generated recruitment content (FH-A5), preferential treatment of internal candidates (FH-A6--FH-A7), recruitment channel selection (FH-A8), and disclosure of screening questions (FH-A9). Other hazards emerge during candidate evaluation and selection, including CV presentation effects (FH-A10), subjective CV review (FH-A11), reviewer fatigue and order effects (FH-A12), external personality assessments (FH-A13), non-standardised interviews (FH-A14), and subjective assessments of candidate attitude (FH-A15).

\begin{table}[t]
\centering
\footnotesize
\caption{Identified fairness hazards in Organisation A hiring process.}
\label{tab:orgA_hazards}
\begin{tabular}{p{0.08\linewidth} p{0.15\linewidth} p{0.62\linewidth}}
\toprule
\textbf{ID} & \textbf{Process Step} & \textbf{Hazard Description} \\
\midrule
FH-A1 & Process oversight &
No actor is explicitly responsible for verifying that all recruitment steps are performed consistently and according to the intended procedure. \\

FH-A2 & Exceptional hiring &
Some candidates may receive opportunities or favourable treatment unavailable to others. \\

FH-A3 & Job description &
Exclusionary or subjective requirements may discourage qualified candidates from applying. \\

FH-A4 & Job description &
Job descriptions may be intentionally or unintentionally tailored to favour a preferred candidate profile. \\

FH-A5 & Generative AI usage &
AI-generated recruitment content may reinforce stereotypes or organisational assumptions. \\

FH-A6 & Internal candidate selection &
Internal candidates may be selected without a comparative evaluation against external candidates when the organisation considers the internal candidate sufficiently qualified and available. \\

FH-A7 & Internal candidate selection &
When multiple internal candidates are available, one candidate may be favoured over another without a transparent or comparable evaluation process. \\

FH-A8 & Recruitment channels &
Selected recruitment channels may provide unequal access to job opportunities. \\

FH-A9 & Screening questions &
Some candidates may obtain prior knowledge of screening questions and gain an unfair advantage during screening. \\

FH-A10 & CV submission &
Differences in CV format or structure may influence evaluations independently of candidate qualifications. \\

FH-A11 & CV review &
Reviewers may favour candidates with familiar backgrounds or characteristics. \\

FH-A12 & CV review &
Candidate evaluations may be influenced by review order and/or fatigue. \\

FH-A13 & External assessment &
Personality assessments may rely on subjective or insufficiently validated criteria. \\

FH-A14 & First interview &
Non-standardised interviews may produce inconsistent candidate evaluations. \\

FH-A15 & Second interview &
Subjective assessment of attitude may result in unfair evaluations. \\

\bottomrule
\end{tabular}
\end{table}

Building upon the identified hazards, we defined mitigation strategies intended to reduce the likelihood and/or impact of unfair outcomes (step D in FHA). 
The resulting mitigation proposals are reported in Table~\ref{tab:orgA_mitigations}. The revised diagram with mitigation strategies is available in the replication package.

\begin{table}[t]
\centering
\footnotesize
\caption{Mitigation strategies proposed for the identified fairness hazards in the Organisation A hiring process.}
\label{tab:orgA_mitigations}
\begin{tabular}{p{0.08\linewidth} p{0.85\linewidth}}
\toprule
\textbf{FHs} & \textbf{Mitigation Strategy} \\
\midrule

FH-A1 &
\textbf{Fairness audit and accountability mechanism.}
Assign responsibility for periodically reviewing recruitment cases and verifying that all fairness-relevant process steps have been completed and documented. Deviations should be recorded and reported for corrective action. \\

FH-A2 &
\textbf{Independent exceptional-hiring review.}
Require an HR representative not involved in the exceptional hiring process to review and document the justification for bypassing the standard recruitment procedure, identifying potential unfair treatment and recommending corrective actions when needed. \\

FH-A3 &
\textbf{Fairness review of job descriptions.}
Require job descriptions to undergo an independent review by HR personnel not involved in the recruitment. The review should identify exclusionary, subjective, or unnecessary requirements. Fairness-support tools, including generative AI systems specifically configured for fairness analysis, may assist reviewers. \\

FH-A4 &
\textbf{Job-description justification and conflict-of-interest checks.}
Require justification for highly specific requirements and maintain documented evidence of their relevance to the position. Introduce conflict-of-interest declarations and organisational rules preventing recruitment decisions involving close personal relationships. \\

FH-A5 &
\textbf{Multi-perspective AI review.}
Review AI-generated recruitment content using alternative prompts, independent reviewers, or additional AI systems before publication. The objective is to identify potentially biased assumptions and overlooked exclusionary language. \\

FH-A6 &
\textbf{Internal-candidate justification.}
When an internal candidate is selected without opening external recruitment, document the reasons for the decision and the criteria used to determine suitability. The justification should be reviewed by HR and archived for accountability purposes. \\

FH-A7 &
\textbf{Transparent internal recruitment process.}
Advertise internal opportunities through a formal internal recruitment procedure, allowing eligible employees to express interest and be evaluated using consistent criteria. \\

FH-A8 &
\textbf{Recruitment-channel diversity policy.}
Define minimum channel-diversity requirements. Positions should be advertised through multiple complementary channels to reduce the risk of systematically excluding specific candidate groups. \\

FH-A9 &
\textbf{Transparent screening criteria.}
Publish the evaluation criteria and rationale used in the screening phase so that all candidates have access to the same information and expectations. \\

FH-A10 &
\textbf{Standardised application format.}
Provide candidates with a standard application template or CV format to reduce the influence of presentation style on candidate evaluation. \\

FH-A11 &
\textbf{Structured CV-evaluation rubric.}
Define explicit evaluation criteria and scoring rules for assessing applications. Reviewers should document the rationale for their evaluations using the rubric. \\

FH-A12 &
\textbf{Balanced review allocation.}
Randomise application order, distribute reviews across multiple sessions, and periodically compare evaluations of similar candidates to identify inconsistencies potentially caused by fatigue or ordering effects. \\

FH-A13 &
\textbf{Structured assessment interpretation.}
Define explicit guidelines for interpreting personality-assessment results and require assessors to justify how assessment outcomes relate to job-relevant competencies. \\

FH-A14 &
\textbf{Structured interview protocol.}
Use a predefined set of questions and evaluation criteria for all candidates applying for the same position while allowing limited location-specific follow-up questions. \\

FH-A15 &
\textbf{Structured attitude-assessment rubric.}
Define observable behavioural indicators and evaluation criteria for assessing attitude, team fit, and related attributes. Interviewers should justify ratings using documented evidence from the interview. \\

\bottomrule
\end{tabular}
\end{table}

\subsubsection{Organisation B} \label{sec:rq1OrgB}
The hazard-identification procedure for Organisation~B (Step~B of FHA) resulted in the fourteen fairness hazards (FH-B\textit{i}) reported in Table~\ref{tab:orgB_hazards}. Of these hazards, $6$ were identified independently by both analysts. The remaining $8$ were initially identified by only one analyst: $3$ by one analyst and $5$ by the other. All analyst-specific hazards were discussed during the consolidation phase and retained after consensus was reached. The identified hazards span the entire recruitment process, affecting approximately 17\% of the distinct process elements in the validated workflow. A first group concerns process governance and role definition, including the absence of explicit responsibility for overseeing the consistent execution of the recruitment process (FH-B1), potential bias in the definition of new professorships (FH-B2), and direct appointment procedures for student assistants (FH-B3)
. Further hazards affect access to employment opportunities and the definition and communication of vacancies, including exclusionary or tailored job descriptions (FH-B4--FH-B5), AI-generated recruitment content (FH-B6), and the selection of recruitment channels (FH-B7). The remaining hazards arise during candidate assessment and selection, including non-standardised application ratings (FH-B8), familiarity bias in application review (FH-B9), biases introduced through AI-generated interview questions (FH-B10), advantages associated with candidates' familiarity with the same language model used by the organisation (FH-B11), non-standardised interview evaluation (FH-B12), unequal influence during committee discussions (FH-B13), and fatigue or order effects during candidate voting and ranking (FH-B14).

\begin{table}[t]
\centering
\footnotesize
\caption{Identified fairness hazards in the Organisation B hiring process.}
\label{tab:orgB_hazards}
\begin{tabular}{p{0.08\linewidth} p{0.24\linewidth} p{0.56\linewidth}}
\toprule
\textbf{ID} & \textbf{Process Step} & \textbf{Hazard Description} \\
\midrule

FH-B1 & Process oversight &
No actor is explicitly responsible for verifying that all recruitment steps are performed consistently and according to the intended procedure. \\

FH-B2 & Professorship definition &
The definition of a new professorship may be influenced by the interests or priorities of the actors identifying the need, potentially disadvantaging other subject areas or candidate profiles. \\

FH-B3 & Student assistant selection &
Student assistants may be selected directly without a public recruitment process, limiting equal access to opportunities. \\


FH-B4 & Job description &
Exclusionary, subjective, or unnecessarily restrictive requirements may discourage qualified candidates from applying. \\

FH-B5 & Job description &
Job descriptions may be intentionally or unintentionally tailored to favour a preferred candidate profile. \\

FH-B6 & Generative AI usage &
AI-generated recruitment content may reinforce stereotypes or organisational assumptions. \\

FH-B7 & Recruitment channels &
Selected publication channels may provide unequal visibility of job opportunities across candidate groups. \\

FH-B8 & Application rating &
Candidates may be evaluated inconsistently because A/B/C rating criteria are not fully standardised across evaluators. \\

FH-B9 & CV/application review &
Reviewers may favour candidates with familiar educational, professional, institutional, or cultural backgrounds. \\

FH-B10 & Interview-question generation with AI &
AI-generated interview questions may introduce unintended biases or emphasise characteristics unrelated to job performance. \\

FH-B11 & Interview-question generation with AI &
Candidates who use the same language model to prepare for the interview may gain an unintended advantage if that model is also used to generate interview questions, as they may become more familiar with its generation patterns, style, and implicit assumptions.\\

FH-B12 & Interview evaluation &
Candidates may be evaluated inconsistently because interview structure and evaluation criteria are not fully standardised. \\

FH-B13 & Committee discussion &
Committee discussions may be disproportionately influenced by dominant personalities or senior members, reducing equal consideration of alternative viewpoints. \\

FH-B14 & Committee voting and ranking &
Candidate rankings may be influenced by fatigue or order effects, disadvantaging candidates evaluated later. \\

\bottomrule
\end{tabular}
\end{table}

The mitigation-planning procedure for Organisation~B (Step~D of FHA) resulted in the strategies reported in Table~\ref{tab:orgB_mitigations}. The proposed mitigations address all identified hazards and include accountability and audit mechanisms (FH-B1), greater transparency and justification in position definition and exceptional recruitment procedures (FH-B2--FH-B3), independent review and standardisation of job descriptions and candidate evaluation criteria (FH-B6--FH-B7, FH-B8--FH-B9, FH-B12), diversified recruitment channels (FH-B7), human oversight and multi-perspective review of AI-supported activities (FH-B6, FH-B10--FH-B11), and safeguards for balanced committee discussion and candidate ranking (FH-B13--FH-B14).

\begin{table}[t]
\centering
\footnotesize
\caption{Mitigation strategies proposed for the identified fairness hazards in the Organisation B hiring process.}
\label{tab:orgB_mitigations}
\begin{tabular}{p{0.08\linewidth} p{0.85\linewidth}}
\toprule
\textbf{FHs} & \textbf{Mitigation Strategy} \\
\midrule

FH-B1 &
\textbf{Fairness audit and accountability mechanism.}
Assign responsibility for periodically reviewing recruitment cases to verify that all fairness-relevant process steps have been completed and documented. Deviations should be recorded and reported for corrective actions. \\

FH-B2 &
\textbf{Justification of position definition.}
Before opening a new professorship, assess whether the proposed disciplinary focus contributes to a balanced coverage of the institution's research and teaching areas. Periodically review the distribution of professorships across disciplines and, where appropriate, benchmark it against comparable universities to identify potential over- or under-representation of specific subject areas. \\

FH-B3 &
\textbf{Transparent student-assistant recruitment.}
Whenever feasible, advertise student-assistant positions internally before direct appointment. When direct recruitment is necessary, document the justification and selection criteria to ensure transparency and accountability. \\

FH-B4 &
\textbf{Fairness review of job descriptions.}
Require job descriptions to undergo an independent review by HR personnel not directly involved in the recruitment. The review should identify exclusionary, subjective, or unnecessarily restrictive requirements. Fairness-support tools, including generative AI systems configured for fairness analysis, may assist reviewers. \\

FH-B5 &
\textbf{Job-description justification.}
Require justification for highly specific requirements and document how each requirement relates to the responsibilities of the position. Additional reviewers should verify that the description does not unintentionally favour a predetermined candidate profile. \\

FH-B6 &
\textbf{Multi-perspective AI review.}
Review AI-generated recruitment content using alternative prompts, independent reviewers, or additional AI systems before publication to identify potentially biased assumptions or exclusionary language. \\

FH-B7 &
\textbf{Recruitment-channel diversity policy.}
Define minimum publication requirements by combining institutional, professional, and public recruitment channels to maximise equal visibility of employment opportunities. \\

FH-B8 &
\textbf{Structured application-rating rubric.}
Define explicit A/B/C evaluation criteria and scoring rules. Reviewers should document the rationale supporting each assigned rating to improve consistency across evaluators. \\

FH-B9 &
\textbf{Structured application-review rubric.}
Provide reviewers with explicit evaluation criteria and require documented justification for candidate assessments to reduce subjective judgement and familiarity bias. \\

FH-B10 &
\textbf{Human review of AI-generated interview questions.}
Require AI-generated interview questions to be reviewed and approved by the interview panel to ensure relevance, fairness, and alignment with the competencies being assessed. \\

FH-B11 &
\textbf{Transparent interview framework.}
Use multiple language models or prompting strategies and periodically revise the question set to reduce model-specific biases and predictable question patterns. \\

FH-B12 &
\textbf{Interview evaluation rubric.}
Develop and use a rubric defining the evaluation criteria, behavioural indicators, and scoring rules for assessing candidates applying for the same position. Interviewers should document the rationale supporting each assigned score. \\

FH-B13 &
\textbf{Moderated committee discussion.}
Assign an independent chair or moderator responsible for ensuring balanced participation during committee discussions and explicitly inviting alternative viewpoints before reaching conclusions. Require committee members to record an initial individual assessment before group discussion or voting. Final rankings should preserve traceability between individual and collective evaluations.\\

FH-B14 &
\textbf{Balanced evaluation scheduling.}
Randomise candidate discussion order when possible, distribute evaluations across multiple sessions, and periodically review rankings to identify inconsistencies potentially caused by fatigue or order effects. \\

\bottomrule
\end{tabular}
\end{table}

\subsubsection{Cross-Case Synthesis}\label{sec:rq1synthesis}
Following the separate analyses of the two recruitment workflows, the identified fairness hazards were compared and grouped based on similarities in their underlying sources and mechanisms. Table~\ref{tab:cross_case_hazard_types} presents the resulting cross-case categorisation and indicates the hazards associated with each category in Organisations~A and~B.

\begin{table}[t]
\centering
\footnotesize
\caption{Cross-case categories of fairness hazards identified in the two processes.}
\label{tab:cross_case_hazard_types}
\begin{tabular}{p{0.22\linewidth} p{0.42\linewidth} p{0.13\linewidth} p{0.13\linewidth}}
\toprule
\textbf{Hazard Category} &
\textbf{Description} &
\textbf{Organisation A} &
\textbf{Organisation B} \\
\midrule

Process-governance gap &
Absence of explicit responsibility for ensuring that recruitment activities and safeguards are performed consistently. &
FH-A1 &
FH-B1 \\

Unequal access to opportunities &
Some individuals may receive privileged access to employment opportunities, while others may be excluded or reached less effectively. &
FH-A2, FH-A6, FH-A7, FH-A8 &
FH-B3, FH-B7 \\

Biased specification of opportunities &
Job descriptions or requirements may be subjective, exclusionary, unnecessarily restrictive, or tailored to favour a preferred profile. &
FH-A3, FH-A4 &
FH-B4, FH-B5 \\

AI-mediated content bias &
AI-generated recruitment content may reproduce stereotypes, organisational assumptions, or criteria unrelated to job performance. &
FH-A5 &
FH-B6, FH-B10 \\

Unequal informational advantage &
Some candidates may possess process-relevant information or familiarity unavailable to others, gaining an advantage unrelated to their qualifications. &
FH-A9 &
FH-B11 \\

Subjective or affinity-based evaluation &
Candidate assessment may be influenced by familiarity, perceived similarity, subjective constructs, or insufficiently validated criteria. &
FH-A11, FH-A13, FH-A15 &
FH-B9 \\

Procedural evaluation inconsistency &
Candidates may be assessed under non-equivalent conditions because rating criteria, interview structures, or evaluation procedures are insufficiently standardised. &
FH-A14 &
FH-B8, FH-B12 \\

Cognitive workload and order effects &
Evaluation may be influenced by fatigue, presentation order, or limitations of human attention. &
FH-A12 &
FH-B14 \\

Representation-format bias &
The format in which candidate information is presented may influence evaluation independently of substantive qualifications. &
FH-A10 &
-- \\

Institutional agenda-setting bias &
The definition of institutional needs may reflect the priorities of particular actors, disadvantaging other subject areas or candidate profiles. &
-- &
FH-B2 \\

Collective decision-making bias &
Committee composition, hierarchical status, or unequal participation may systematically influence collective candidate evaluation. &
-- &
FH-B13 \\

\bottomrule
\end{tabular}
\end{table}

The cross-case comparison identified \textit{eight} recurring hazard categories across the two organisations. These concerned: process governance, access to employment opportunities, the specification of job requirements, AI-generated content, unequal informational advantages, subjective evaluation, procedural inconsistency, and cognitive workload or order effects. Their recurrence across organisationally different recruitment processes suggests that they are associated with common characteristics of hiring workflows rather than with a single organisational setting.

Three categories were context-specific. Representation-format bias was identified only in Organisation~A, where the presentation of candidate information could influence evaluation. Institutional agenda-setting bias and collective decision-making bias were specific to Organisation~B and reflected characteristics of the higher-education context, including the definition of professorships and the formal role of selection committees. Overall, the findings indicate that the majority of hazard categories recur across the two recruitment processes, while some emerge from organisation-specific governance structures and decision-making practices.

The mitigation strategies proposed for the two recruitment processes were compared based on the control mechanisms they use to prevent, detect, or reduce unfairness. Table~\ref{tab:cross_case_mitigation_types} presents the resulting cross-case categorisation. The categories are not necessarily mutually exclusive, since a mitigation strategy may serve more than one purpose. For example, the review of AI-generated content may combine independent verification with AI-specific oversight.

\begin{table}[t]
\centering
\footnotesize
\caption{Cross-case categories of mitigation strategies proposed for the two
recruitment processes.}
\label{tab:cross_case_mitigation_types}
\begin{tabular}{
p{0.23\linewidth}
p{0.43\linewidth}
p{0.12\linewidth}
p{0.12\linewidth}
}
\toprule
\textbf{Mitigation Category}
& \textbf{Description}
& \textbf{Organisation A}
& \textbf{Organisation B} \\
\midrule

Governance and accountability
& Assign explicit responsibilities, introduce oversight and audit mechanisms,
document deviations, and require justification for exceptional or
consequential decisions.
& FH-A1, FH-A2, FH-A6
& FH-B1--FH-B3 \\

Independent review and integrity safeguards
& Require independent verification, separation of responsibilities,
documented justification, or conflict-of-interest safeguards before
recruitment artefacts or decisions are finalised.
& FH-A3--FH-A5
& FH-B4--FH-B7, FH-B11 \\

Transparency and equal access
& Ensure that opportunities, selection criteria, and process-relevant
information are accessible on equivalent terms to potentially eligible
candidates.
& FH-A7--FH-A9
& FH-B3, FH-B8 \\

Standardisation and structured evaluation
& Reduce variability and subjectivity through standardised artefacts,
rubrics, predefined criteria, scoring rules, and documented evaluation
rationales.
& FH-A10, FH-A11, FH-A13--FH-A15
& FH-B9, FH-B10, FH-B13 \\

AI oversight and diversification
& Introduce human review and reduce dependence on a single AI system, model,
prompting strategy, or generated output.
& FH-A5
& FH-B7, FH-B11--FH-B12 \\

Collective decision-process safeguards
& Structure collective decisions through composition guidelines, balanced
participation, independent initial assessments, moderated discussion, and
traceable deliberation.
& --
& FH-B4, FH-B14 \\

Workload and order-effect management
& Reduce fatigue, ordering effects, and other cognitive biases through
balanced workload allocation, scheduling, or changes in evaluation order.
& FH-A12
& FH-B15 \\

\bottomrule
\end{tabular}
\end{table}

Most mitigation categories were represented in both organisational cases. Recurring strategies concerned governance and accountability, independent review, transparency and equal access, standardisation of evaluation, AI oversight, and workload management. This similarity reflects the recurrence of the corresponding hazard mechanisms across the two recruitment processes. Although the specific controls differed, both analyses frequently recommended making responsibilities explicit, introducing additional review, standardising consequential evaluations, and documenting the rationale for decisions.

Collective decision-process safeguards were identified as a distinct category only in Organisation~B, where formal selection committees played a central role in candidate evaluation and ranking. This category reflects the need to address not only individual evaluator bias but also power imbalances, hierarchical influence, and unequal participation in deliberative bodies.

Overall, the mitigation strategies combine preventive, detective, and corrective controls. Some seek to prevent hazards by standardising procedures or broadening access, others introduce review mechanisms capable of detecting potential unfairness before a decision is finalised, and others support the revision or justification of decisions. The findings therefore suggest that fairness mitigation in recruitment cannot be reduced to technical interventions, but requires coordinated organisational, procedural, human, and technological controls.

\subsection{RQ2: Practitioner Perceptions of FHA}
The thematic analysis of the organisational feedback identified three themes: (i) the context-dependent validation and completeness of the identified hazards; (ii) the feasibility, adaptation, and organisational fit of the
proposed mitigations; and (iii) the reflective usefulness and
awareness-raising potential of FHA. The themes capture patterns shared across the two cases while retaining differences associated with their organisational
contexts.

\begin{table}[t]
\centering
\footnotesize
\caption{Cross-case themes emerging from the organisational feedback.}
\label{tab:rq2_themes}
\begin{tabular}{p{0.25\linewidth} p{0.38\linewidth} p{0.29\linewidth}}
\toprule
\textbf{Theme} &
\textbf{Cross-case finding} &
\textbf{Main difference between cases} \\
\midrule

Context-dependent validation and completeness of hazards &
Practitioners assessed the realism of the hazards in relation to actual organisational practices, existing safeguards, and the conditions under which each hazard could arise. &
Organisation~A mainly confirmed the identified hazards, whereas
Organisation~B more frequently corrected them based on formal procedures and existing controls. \\

Feasibility, adaptation, and organisational fit of mitigations &
Most mitigations were regarded as feasible or useful, but their implementation depended on organisational fit, proportionality to the risk, resources, and existing
practices. &
Organisation~A emphasised candidate experience and organisational identity;
Organisation~B more often identified mitigations that were already partly or fully implemented and highlighted procedural and scheduling constraints. \\

Reflective usefulness and awareness-raising potential of FHA &
In both organisations, FHA encouraged reflection on underexamined risks and made the fairness relevance of organisational practices more explicit. &
Organisation~A particularly highlighted fairness concerns related to generative AI, whereas Organisation~B also recognised existing practices as implicit fairness safeguards. \\

\bottomrule
\end{tabular}
\end{table}

\smallskip
\textbf{\textit{Context-Dependent Validation and Completeness of Hazards.}}
Across both cases, the realism of a hazard was not assessed as a purely binary property. Rather, participants interpreted each hazard in relation to the actual workflow, the conditions under which the issue could arise, and the safeguards already in place within the organisation. \textbf{Organisational knowledge was therefore necessary to distinguish hazards that directly reflected current practice from those that required qualification or reformulation.}

In Organisation~A, most hazards were accepted as realistic. Qualifications mainly concerned the circumstances under which a risk would arise. For example, the participant distinguished between an exclusionary preference in a job description and a characteristic that was genuinely necessary for the role (\textit{``If the job description requires candidates to be extroverted, then a person who is not extroverted is not qualified.''}). Similarly, the risk associated with external personality assessments was considered possible, but less salient because the organisation relied on professional external providers.

In Organisation~B, evaluations were more frequently shaped by formal
governance arrangements and existing safeguards. Some hazards were recognised as potential risks but were considered to have already been reduced through strategic approval procedures, multiple reviewers, explicit process ownership, or mandatory documentation. The process-oversight hazard, for instance, was rejected as formulated because responsibility for supervising recruitment and approval gates was already explicitly assigned (\textit{``For professorships, I’m in charge for the whole process. If it’s not done properly, you do not get the signatures and approval [and] will not be able to finalise the procedure.''}).

The two cases also differed in completeness. The participant from Organisation~A considered that no important fairness risks had been omitted.
In Organisation~B, by contrast, participants identified an additional concern involving informal or exceptional personnel situations that fall outside the formally modelled recruitment workflow, referring to the \textit{``messy world around these processes''}. This observation indicates that the boundaries of the process model can themselves influence which fairness hazards become visible.

\smallskip
\textbf{\textit{Feasibility, Adaptation, and Organisational Fit of Mitigations.}}
Most proposed mitigations were evaluated positively in both organisations. However, participants did not treat them as controls that could be adopted independently of their organisational context. \textbf{Their feasibility depended on whether they were proportionate to the risk, compatible with current practices, and sustainable in terms of resources and workload.}

In Organisation~A, independent review, explicit evaluation criteria, conflict-of-interest controls, channel diversification, and the review of AI-generated content were generally regarded as feasible. The clearest exception concerned the proposal to require a standardised application format. Although the associated hazard was considered realistic, the mitigation was viewed as inconsistent with the organisation's personalised recruitment approach (\textit{``Very large companies can do it. But we are much more personal. I wouldn’t do it. A standard application template [means] you’re already policing in the first encounter.''}). As the participant explained, the hazard was \textit{``super realistic''}, but the mitigation would not currently work in the organisation. This distinction shows that acceptance of a hazard does not necessarily entail acceptance of the proposed response.

In Organisation~B, many mitigations corresponded to safeguards already fully or partially implemented, including strategic justification, multiple reviewers, committee discussion, breaks between interviews, and formal approval procedures. Participants nevertheless suggested extensions, such as making evaluation rationales mandatory, involving diversity personnel in the choice of recruitment channels, comparing outputs from alternative AI models, and collecting individual assessments before committee discussion.
One participant particularly endorsed the latter mechanism because it could make the perspectives of \textit{``silent voices''} visible before collective deliberation.

The barriers identified in the two cases also differed. Organisation~A primarily emphasised cost and candidate burden. Organisation~B additionally highlighted access to specialised recruitment networks, informal and individually managed practices, the difficulty of scheduling large committees, and the limits of using detailed criteria to standardise inherently qualitative judgements. Overall, the feedback suggests that mitigation strategies should be treated as adaptable organisational safeguards rather than as controls to be transferred unchanged across contexts.

\smallskip
\textbf{\textit{Reflective Usefulness and Awareness-Raising Potential of FHA.}}
Feedback from both organisations indicates that \textbf{FHA was valued not only for producing a list of hazards and mitigations, but also for encouraging reflection on fairness implications that had not previously been examined systematically.}

In Organisation~A, the participant stated that the analysis could improve fairness awareness within the human-resources function and described the session as a beneficial exercise (\textit{``You brought some topics to the surface that I need to think about''}). 

Organisation~B participants similarly described the analysis as providing a new perspective on their recruitment practices. The discussion prompted concrete considerations regarding alternative AI models, recruitment channels for underrepresented groups, and mechanisms to ensure that less vocal committee members could contribute to collective decisions. FHA also helped participants recognise that several routine practices already served as fairness safeguards, even though they had not previously been interpreted in those terms. As one participant observed, many mitigation actions were already performed in everyday practice, \textit{``but we are not aware of this''}.

Across the two cases, FHA therefore supported two complementary forms of reflection: it surfaced fairness concerns that had received limited prior attention and made the fairness function of existing organisational practices more explicit.

\section{Discussion}\label{sec:discussion}

\subsection{Integrated Interpretation of the Cross-Case Findings}
\label{sec:integratedDiscussion}

The findings from RQ1 and RQ2 provide complementary perspectives on the
applicability of FHA in real organisational environments. RQ1 showed that the fairness hazards identified in the two recruitment workflows reflected both recurring process characteristics and organisation-specific arrangements.
RQ2, in turn, showed that practitioners assessed the realism of those hazards in relation to the actual organisational context, including existing safeguards, exceptional circumstances, legitimate role requirements, and the boundaries of the modelled workflow. Taken together, these findings indicate that \textbf{the contextual character of fairness hazards concerns not only their identification, but also their subsequent validation and interpretation}.

Several hazard categories appeared in both organisations despite differences in governance, decision-making structures, and recruitment procedures. This recurrence suggests that some fairness concerns arise from common features of recruitment, such as defining and communicating opportunities, progressively screening candidates, applying human judgement, and coordinating multiple actors and supporting technologies. Nevertheless, similar underlying concerns, such as unequal access or inconsistent evaluation, were produced by different procedures, exceptions, and decision mechanisms in the two organisations.
Practitioners' feedback further showed that an analytically plausible hazard may be less salient, apply only under specific conditions, or already be partly controlled by organisational practices. \textbf{The realism of a hazard should therefore not be treated as a simple binary property, but as a judgement that must be grounded in knowledge of the actual process}.

Other hazards were closely tied to the organisational context.
The feedback concerning completeness reinforced this contextual dependence. Organisation~A considered that no important risks had been omitted, whereas Organisation~B identified an additional concern involving informal or exceptional personnel situations outside the formally modelled recruitment workflow. This finding highlights the \textbf{importance of examining not only the activities represented in the process model, but also exceptions, informal practices, and interactions occurring around the formal process}. A reusable catalogue of known recruitment hazards may therefore support analysis, but cannot replace detailed examination and organisational validation of the actual workflow.

A particularly relevant cross-case finding concerns the relationship between hazards and mitigations. \textbf{While the hazards identified through the FHA were often context-dependent, the proposed mitigation strategies converged on a smaller, more stable set of organisational interventions}. Across the two cases, recurrent mitigation types included explicit accountability, independent review, transparency and equal access, structured evaluation, human oversight of AI-supported activities, safeguards for collective deliberation, and workload management. Different hazards could therefore be addressed through similar control mechanisms, adapted to the process step and organisational setting in which they were introduced.

Practitioners generally regarded the proposed mitigations as feasible or useful, but their assessments also qualified their transferability. Some mitigations corresponded to safeguards that were already fully or partly
implemented, particularly in Organisation~B. Others required adaptation to existing procedures or organisational constraints. In Organisation~A, for example, a standard application format was considered unsuitable despite the associated hazard being deemed realistic, because the proposed control would increase the candidate burden and conflict with the organisation's personalised recruitment approach. Organisation~B instead highlighted constraints associated with cost, access to specialised recruitment networks, the scheduling of large committees, and the limits of standardising qualitative human judgement. These findings show that acceptance of a hazard does not automatically imply acceptance of the proposed mitigation.

\textbf{Hazards and mitigations consequently exhibit different, but not absolute, degrees of transferability.} Fairness hazards must be elicited and validated in relation to the specific workflow, whereas mitigation strategies may be organised as reusable fairness-by-design patterns. Such patterns could provide analysts with a preliminary repertoire of organisational and procedural interventions. Their reuse, nevertheless, requires consideration of implementation responsibilities, existing safeguards, proportionality, resource constraints, and potential effects on the actors involved in the process.

The findings from RQ2 also indicate that FHA may have a reflective function beyond the production of hazard and mitigation tables. Participants in both organisations reported that the analysis brought underexamined fairness issues to their attention. In both organisations, this reflection particularly concerned the use of generative AI in recruitment, and prompted consideration of alternative AI tools. It also helped participants recognise that several routine practices already functioned as fairness safeguards, although they had not previously been interpreted in those terms. \textbf{FHA can therefore support both the identification of previously underexamined risks and the explicit recognition of fairness-relevant controls already embedded in organisational practice}.

Overall, the integrated findings suggest that the applicability of FHA does not derive from producing universally valid lists of hazards and mitigations. Rather, it derives from providing a structured basis for examining a concrete workflow, confronting the resulting analysis with organisational knowledge, and adapting possible interventions to the conditions of the organisation.
The hazard and mitigation categories resulting from this study should therefore be regarded as a preliminary analytical resource rather than an exhaustive classification.

\subsection{Implications for Fairness Hazard Analysis}
\label{sec:fhaImplications}

\textbf{The findings have several implications for the application and further development of FHA, particularly regarding the composition of the analysis team}. Organisational members with direct knowledge of the process should play a central role throughout the analysis, rather than being involved only after hazards and mitigations have been identified. Their knowledge is essential for understanding how the process is actually performed, under which conditions deviations occur, which safeguards are already in place, and whether an analytically plausible hazard accurately reflects organisational practice.
At the same time, fairness and methodological expertise remain necessary for structuring the analysis, challenging assumptions, and identifying potential sources of unfairness. Analysts who are less familiar with the application domain may also question practices that insiders take for granted, similarly to the role of the ``smart ignoramus'' described by Berry~\cite{berry2002importance}. FHA should therefore be applied by a multidisciplinary team that combines detailed organisational and process knowledge with fairness and methodological expertise. Such complementarity may broaden the range of hazards considered, although further studies are needed to investigate how the number and composition of analysts affect hazard identification.

The findings also indicate that \textbf{Step~A of FHA should elicit more than the nominal sequence of process activities.} The system definition should explicitly capture existing safeguards, responsibility allocations, approval mechanisms, and other controls that may already reduce the likelihood or consequences of fairness hazards. When such controls are omitted from the process description, the subsequent analysis may propose mitigations that are already implemented or incorrectly represent a residual risk as an unmitigated hazard. Elicitation should therefore include targeted questions about how decisions are reviewed, who is responsible for verifying compliance with the process, which approvals
are required, and which formal or informal safeguards operate at each relevant process step.

Step~A should likewise examine deviations from the standard workflow, including exceptional paths, ad hoc arrangements, and situations for which no established procedure exists. In Organisation~A, exceptional recruitment situations were elicited and represented as part of the workflow, allowing the analysis of their fairness implications directly. In Organisation~B, however, the feedback session revealed a further concern involving personnel situations outside the formally modelled recruitment process. This indicates that process elicitation should explicitly probe not only documented exceptions, but also informal practices and cases handled outside ordinary procedures. Strengthening Step~A in this way would provide a more complete basis for the subsequent identification and assessment of fairness hazards.

\textbf{Mitigation planning should also distinguish between the objective of a control and its specific implementation.} The feedback showed that practitioners could agree with the fairness objective of a mitigation while rejecting its proposed form. Mitigations should therefore be formulated as adaptable intervention patterns rather than prescriptive solutions. For example, standardisation may be achieved through a mandatory template, a structured review rubric, or automated extraction of comparable information, depending on the process and organisational constraints. FHA should support the comparison of alternative implementations that pursue the same fairness objective.

\subsection{Implications for Requirements Engineering and Practice}

From an RE perspective, FHA provides a risk-oriented mechanism for deriving fairness-relevant organisational and system requirements. The proposed mitigations can be translated into requirements concerning, for example, responsibility allocation, independent review, justification of decisions, traceability of AI-generated content, disclosure of evaluation criteria, multi-person assessment, documentation of individual judgements, and workload controls. \textbf{This creates a traceable relationship between an identified fairness hazard, the decision point at which its effects may be observed, and the organisational or technical requirement introduced to control it}.

The findings also suggest that existing \textbf{organisational procedures may contain implicit fairness requirements, although their fairness function was not always explicit.}
Making this function visible can help organisations assess whether existing
controls adequately address a hazard, identify gaps between intended and actual practice, and preserve relevant safeguards when processes or supporting technologies change.

\textbf{For practitioners, the emerging mitigation categories may provide a useful starting point for analysing recruitment processes.} They should not be applied as a checklist of mandatory solutions, but as prompts for considering possible controls and their alternatives. The findings indicate that implementation decisions should account for organisational resources, candidate burden, decision-making structures, legal and procedural constraints, and the interaction between formal and informal practices. Practitioner involvement is therefore necessary not only for validating hazards, but also for determining how fairness objectives can be realised without introducing disproportionate or contextually unsuitable controls.

Finally, \textbf{the awareness-raising findings indicate that FHA may support organisational learning.} By connecting familiar procedures with explicit fairness objectives, the analysis can help practitioners articulate why particular controls matter and identify where apparently routine decisions may produce unequal effects. This reflective function may be especially relevant when new technologies, such as generative AI, are introduced into established workflows without a corresponding review of their fairness implications.

\subsection{Progression from TRL 3 to TRL 5}
The proof-of-concept evaluation examined FHA at TRL~3 through expert focus
groups applying and discussing the methodology using a constructed and
simplified, yet realistic, recruitment exemplar. That evaluation provided
initial evidence that the underlying concepts and analytical steps were
understandable and capable of supporting the identification of fairness
hazards and mitigations. It also generated methodological recommendations for applying FHA beyond the constructed exemplar (cfr. Table~\ref{tab:themesFHA}).

The operationalisation adopted in the multiple-case study addressed five of these recommendations, concerning the contextualisation of fairness, the exploration of potentially overlooked hazards, the formulation of mitigation strategies, and the involvement of analysts and organisational participants throughout the analysis.

First, FHA addressed the recommendation to \emph{define fairness explicitly for each analysis context} by interpreting fairness in relation to the recruitment procedures, objectives, applicable principles, and affected stakeholders of each organisation. This supported the distinction between unacceptable bias and differentiation that could be considered intended or justified within the specific process. 
The recommendation to \emph{support the exploration of hidden biases} was addressed by combining complementary perspectives. The two analysts independently examined each workflow and consolidated their findings, while organisational participants contributed detailed knowledge of the process, existing safeguards, and operational context.
Additionally, the recommendation to \emph{encourage iterative, dialogic reflection} was addressed through interaction both between the analysts and with organisational participants. Participants contributed to workflow elicitation and clarification and subsequently assessed the identified hazards and mitigations, enabling the analysis to be refined in light of organisational knowledge. 
Finally, the recommendation to \emph{frame bias as a learning opportunity} was addressed through the feedback sessions. Participants were encouraged to reflect on previously underexamined assumptions, practices, and vulnerabilities, as well as on existing organisational controls that could be recognised as fairness safeguards.

The multiple-case study moved beyond demonstration on a constructed exemplar by applying FHA to two real organisational recruitment workflows characterised by actual procedures, exceptions, governance structures, and existing safeguards.

The integrated evidence from RQ1 and RQ2 supports the assessment of FHA at TRL~5. FHA could be operationalised on real organisational workflows, identify both recurring and context-specific hazards, generate mitigation strategies that practitioners generally regarded as relevant, and support reflection on fairness risks and existing safeguards. Practitioner feedback also revealed where the analysis required correction, contextual qualification, or adaptation, demonstrating that FHA could incorporate organisational knowledge rather than merely impose externally generated conclusions.

\section{Lessons Learned}\label{sec:lessons}
The application of FHA to the two organisational workflows yielded the
following methodological and practical lessons.

\begin{enumerate}
\item \textbf{Conceptual bias frameworks are useful as prompts, but cannot
    replace contextual process knowledge.}
    The root causes proposed in the Fairness Debt literature \cite{de2025software} provided useful sensitising concepts for reasoning about possible sources of unfairness, but they did not cover all the hazards emerging from the organisational recruitment workflows. 
    Fairness-related taxonomies should therefore be used as analytical prompts rather than as exhaustive hazard catalogues. Detailed knowledge of the process, contributed by organisational members with different roles and perspectives, remains necessary for identifying context-specific hazards.

    \item \textbf{The composition of the analysis team affects the range of
    hazards that can be identified.}
    In this study, hazard identification benefited from complementary forms of expertise in the analysis team, including knowledge of fairness and bias, empirical-study design, threats to validity, and recruitment practices. Familiarity with academic recruitment was particularly relevant for interpreting the workflow of Organisation~B, both from the perspective of candidates and of people participating in selection procedures. FHA should therefore be applied by a multidisciplinary team combining organisational process knowledge with fairness and methodological expertise.

    Domain familiarity should nevertheless be complemented by an informed
    outsider perspective. An analyst who is not accustomed to the domain may question practices and assumptions that organisational members consider self-evident, similarly to the role of the ``smart ignoramus'' discussed by Berry~\cite{berry2002importance}. Future studies should investigate how the number and profiles of analysts affect the number and types of hazards identified, and whether the marginal contribution of additional analysts eventually decreases. Nielsen and Landauer's model of problem discovery in usability evaluation~\cite{nielsen1993mathematical} may provide an analogy for designing such an investigation.

    \item \textbf{Fairness analysis must distinguish unfair bias from justified differentiation and constrained residual risk.}
    Some requirements or selection decisions may be justified by the
    responsibilities of a position, legal obligations, or legitimate
    organisational objectives. In other situations, organisational or external constraints may prevent a fairness risk from being completely eliminated.
    FHA should therefore distinguish among potentially unfair differentiation, contextually justified differentiation, and residual fairness risks that remain after feasible controls have been considered. Such judgements should be explicitly justified and documented rather than describing a bias as simply ``inevitable''.

    \item \textbf{Mitigation planning requires direct involvement from process owners and organisational practitioners.}
    Analysts who are not involved in the process may identify the objective of a mitigation but may be unable to determine whether its proposed implementation is feasible. The organisational feedback showed that feasibility depends on costs, responsibility allocation, existing procedures, candidate burden, and organisational identity. Organisational members should therefore participate directly in mitigation planning rather than only validating mitigations formulated externally.

    Where comprehensive structural interventions cannot be introduced
    immediately, lower-cost measures---such as documented rationales,
    declarations of potential conflicts of interest, review checkpoints, and awareness activities---may provide incremental safeguards. These measures should not, however, be treated as substitutes for stronger interventions when the severity of the hazard requires them.

    \item \textbf{Visual and spatial representations supported systematic
    workflow inspection.}
    Representing each workflow through a simplified activity diagram provided the analysts with a common visual reference for following branches, decision points, actors, and exceptional paths. Annotating printed diagrams by hand also made the spatial distribution of the identified hazards visible and helped the analysts notice process regions that had received comparatively little attention. This trace was less immediately visible when hazards were represented only in tabular form.

    This observation concerns the analysts' experience rather than a formal
    evaluation of the notation. The study does not demonstrate that UML
    activity diagrams are inherently understandable to practitioners without modelling expertise, nor that paper-based annotation is superior to alternative representations. These aspects should be examined empirically.

    \item \textbf{Recurring hazard categories support reuse, but not
    context-free application.}
    Several hazard categories recurred across the two workflows despite
    differences between the organisations and their recruitment procedures.
    This recurrence suggests that a preliminary library of recruitment-related fairness hazards could support future FHA applications.

    Such a library should be used to stimulate investigation rather than as a checklist assumed to be complete. 

    \item \textbf{AI can function both as a source of fairness hazards and a component of their mitigation.}
    In the constructed proof-of-concept exemplar, AI-supported recruitment was a prominent source of potential hazards. In the real workflows, AI had a more varied role. Repeated reliance on the same model or prompting approach could reproduce similar assumptions or stereotypes, whereas alternative  models and fairness-oriented prompts were proposed as mechanisms for reviewing recruitment content and introducing additional perspectives.

    AI-supported review should nevertheless remain subject to human judgement, traceability, and comparison with other sources of evidence. Research on human--AI feedback loops shows that exposure to biased AI judgements can influence and amplify human bias~\cite{glickman2025human}; it does not imply
    that merely reusing the same prompt causes a model to learn progressively
    from the organisation. FHA should therefore examine both the immediate
    outputs of AI-supported activities and how repeated human reliance on those
    outputs may shape subsequent organisational decisions.
\end{enumerate}

\section{Threats to Validity}\label{sec:threats}

\smallskip
\textbf{\textit{Construct Validity.}}
Fairness hazards are context-dependent analytical constructs rather than directly observable properties of a process. Their identification and formulation may therefore reflect the analysts' interpretation of fairness, the available process information, and the assumptions adopted during the analysis. To mitigate this threat, hazard identification has been performed by two independent analysts, and organisational participants have assessed whether the hazards accurately reflect their practices and whether relevant risks have been omitted.  Similarly, practitioners evaluated the organisational relevance and feasibility of the proposed mitigations, but they were not implemented or evaluated for their effects on recruitment decisions. The findings therefore concern the applicability of FHA and the perceived relevance of its outputs, rather than the effectiveness of the proposed controls.

\smallskip
\textbf{\textit{Internal Validity.}}
The reconstructed workflows may have omitted activities, safeguards, exceptional paths, or informal practices discussed insufficiently during elicitation. Organisational participants reviewed and clarified the workflows before FHA was applied, reducing the likelihood that hazards would be derived from inaccurate process representations. However, the feedback session in Organisation~B revealed both an existing control that had not been captured correctly and an informal situation outside the modelled workflow. This shows that participant validation reduces, but does not eliminate, the risk of an incomplete system definition.

A further threat arises from the involvement of researchers who contributed to the development of FHA and may therefore have been inclined to identify findings supporting its usefulness. Two researchers initially analysed each workflow independently and subsequently reached consensus, while organisational participants could reject, qualify, or extend the resulting analysis. Even with these measures, researcher expectations may have affected hazard formulation and interpretation. Participants may also have been inclined to provide favourable assessments because they had already contributed to workflow elicitation and validation. Moreover, the feedback reflects the perspectives of personnel responsible for the processes and does not include candidates or other stakeholders potentially affected by recruitment decisions.

\smallskip
\textbf{\textit{External Validity}}
The study involved two organisations selected opportunistically on the basis of access and availability, both within the recruitment domain. The cases are not statistically representative of organisations or recruitment processes,  and the findings should not be generalised through statistical inference.
Instead, the study supports case-based analytical generalisation \cite{wieringa2015six}. The recurring hazards and mitigation mechanisms are treated as analytical propositions concerning components of socio-technical processes that may vary less than the organisational cases as a whole.

The findings are most plausibly transferable to recruitment and other bureaucratic or decision-making processes characterised by formal procedures, multiple organisational actors, substantial human judgement, exceptional paths, and interactions with software or AI-supported tools. Transferability to less formal processes, different legal or cultural environments, or settings in which decisions are predominantly automated remains uncertain.
The recurring hazard and mitigation categories should therefore be regarded as preliminary resources whose relevance must be examined against the workflow and organisational context of each new application. Further applications in different domains and organisations are needed to refine their scope.

\section{Conclusion}\label{sec:conclusion}

This paper extends Fairness Hazard Analysis (FHA), a structured methodology for identifying, analysing, and mitigating fairness hazards in socio-technical processes at the requirements level. Building on its proof-of-concept validation, we refined FHA for real organisational workflows and evaluated it through a qualitative multiple-case study of two recruitment processes. The study advances FHA from TRL~3 to TRL~5 and operationalises it through workflow elicitation and modelling, independent hazard analysis, and iterative validation with practitioners.

The findings show that FHA identifies potential biases arising not only from software and AI technologies, but also from organisational procedures, human judgement, and interaction between process activities and roles. The cross-case analysis revealed both recurring patterns and organisation-specific risks, suggesting that reusable knowledge can support, but not replace, contextual analysis.

Practitioners considered the identified hazards largely relevant and most mitigations appropriate, while noting that feasibility depends on available resources to implement mitigation strategies. FHA also encouraged reflection by surfacing underexamined fairness concerns and clarifying the fairness-related role of existing practices.

The results further suggest that FHA should be applied collaboratively and iteratively, combining expertise in fairness, RE, and the application domain. Recurring mitigation strategies identified during the experience—--such as independent review, explicit decision criteria, documented rationales, accountability mechanisms, and collective decision-making—--offer preliminary patterns for future applications, provided they are adapted to context.

The study evaluates the applicability of FHA and the perceived relevance of its outputs, but does not establish whether the proposed mitigations improve actual fairness outcomes. Future work should therefore implement and longitudinally evaluate FHA-derived mitigations, involve affected stakeholders, and apply the methodology across further organisations, domains, and regulatory contexts. Additional research should examine analyst-team composition, develop empirical hazard and mitigation libraries, and strengthen guidance for impact and likelihood assessment, which was outside the scope of this evaluation. We are also developing HumaInFlow, an agentic tool intended to support the application of FHA and the simulation of hazard propagation in socio-technical processes.


\backmatter




\bmhead{Declaration of AI and AI-assisted Technologies in the Writing Process} During the preparation of this work, the author(s) used ChatGPT to improve language and support paper revision. After using this tool, the author(s) reviewed and edited the content as needed and take(s) full responsibility for the content of the publication.

\bmhead{Acknowledgements}
We thank the participants from both organisations for their time and valuable contributions to this study.






\bibliography{biblio}

\end{document}